\documentclass[prd,aps, onecolumn,superscriptaddress,longbibliography,nofootinbib, letterpaper]{revtex4-1}

\usepackage{amsmath,graphicx,epsfig}
\usepackage{epstopdf}
\usepackage{euscript}
\usepackage{amsfonts}
\usepackage{amssymb}
\usepackage{float}
\usepackage{tabularx}
\usepackage{tablefootnote}
\usepackage{xcolor}
\usepackage{multirow}
\usepackage{enumitem}
\usepackage{comment}
\usepackage{hyperref}
\hypersetup{
    colorlinks=true,
    linkcolor=blue,
    }
\date{\today}

\usepackage{natbib}
\usepackage{graphicx}

\def\prn#1{{\left(#1\right)}}

\newcommand{\Bp}{\mathcal{B}_{\text{p}}}  
\newcommand{\Bpk}{\mathcal{\tilde{B}}_{\text{p},jk}}

\usepackage[arrows]{ezedits}
\defineEdit{SK}{\color[rgb]{1.0,0.0,0.0}}{\color[rgb]{1.0,0.0,0.0}}
\defineEdit{IAS}{\color[rgb]{0,0.7,0.5}}{\color[rgb]{0,0.7,0.5}}
\defineEdit{PH}{\color[rgb]{0,0.8,0.8}}{\color[rgb]{0.,0.8,0.8}}
\defineEdit{DK}{\color[rgb]{0.9,0.1,0.9}}{\color[rgb]{0.9,0.1,0.9}}
\defineEdit{DB}{\color[rgb]{0.4,0.4,0.9}}{\color[rgb]{0.4,0.4,0.9}}

\begin{document}
\title{A multi-messenger search for exotic field emission with a global magnetometer network}

\author{Sami S. Khamis}
\email{skhamis@physics.ucla.edu}
\affiliation{Department of Physics and Astronomy, University of California, Los Angeles, California 90095, USA}

\author{Ibrahim A. Sulai}
\affiliation{Department of Physics and Astronomy, Bucknell University, Lewisburg, Pennsylvania 17837, USA}

\author{Paul Hamilton}
\affiliation{Department of Physics and Astronomy, University of California, Los Angeles, California 90095, USA}

\author{S. Afach}
\affiliation{Helmholtz Institute Mainz, 55099 Mainz, Germany}
\affiliation{GSI Helmholtzzentrum für Schwerionenforschung GmbH, 64291 Darmstadt, Germany}
\affiliation{Johannes Gutenberg-Universit{\"a}t Mainz, 55128 Mainz, Germany}

\author{B. C. Buchler}
\affiliation{Research School of Physics, Australian National University, Canberra ACT 2601, Australia}

\author{D. Budker}
\affiliation{Helmholtz Institute Mainz, 55099 Mainz, Germany}
\affiliation{GSI Helmholtzzentrum für Schwerionenforschung GmbH, 64291 Darmstadt, Germany}
\affiliation{Johannes Gutenberg-Universit{\"a}t Mainz, 55128 Mainz, Germany}
\affiliation{Department of Physics, University of California, Berkeley, California 94720, USA}

\author{N. L. Figueroa}
\affiliation{Helmholtz Institute Mainz, 55099 Mainz, Germany}
\affiliation{GSI Helmholtzzentrum für Schwerionenforschung GmbH, 64291 Darmstadt, Germany}
\affiliation{Johannes Gutenberg-Universit{\"a}t Mainz, 55128 Mainz, Germany}

\author{R. Folman}
\affiliation{Department of Physics, Ben-Gurion University of the Negev, Be’er Sheva 84105, Israel}

\author{D. Gavilán-Martín}
\affiliation{Helmholtz Institute Mainz, 55099 Mainz, Germany}
\affiliation{GSI Helmholtzzentrum für Schwerionenforschung GmbH, 64291 Darmstadt, Germany}
\affiliation{Johannes Gutenberg-Universit{\"a}t Mainz, 55128 Mainz, Germany}

\author{M. Givon}
\affiliation{Department of Physics, Ben-Gurion University of the Negev, Be’er Sheva 84105, Israel}

\author{Z. D. Gruji\'{c}}
\affiliation{Institute of Physic Belgrade, University of Belgrade, 11080 Belgrade, Serbia}

\author{H. Guo}
\affiliation{State Key Laboratory of Advanced Optical Communication Systems and Networks, Department of Electronics, and Center for Quantum Information Technology, Peking University, Beijing 100871, China}

\author{M. P. Hedges}
\affiliation{Centre for Quantum Computation and Communication Technology, Research School of Physics, The Australian National University, Acton 2601, Australia}

\author{D. F. Jackson Kimball}
\affiliation{Department of Physics, California State University -- East Bay, Hayward, California 94542-3084, USA}

\author{D. Kim}
\thanks{present address: Mechatronics Research, Samsung Electronics, Hwaseong, 18448, South Korea}
\affiliation{Center for Axion and Precision Physics Research, IBS, Daejeon 34051, Republic of Korea}
\affiliation{Department of Physics, KAIST, Daejeon 34141, Republic of Korea}

\author{E. Klinger}
\affiliation{Universit\'e Marie et Louis Pasteur, SUPMICROTECH, CNRS, Institut FEMTO-ST, F-25000 Besan\c{c}on, France}

\author{T. Kornack}
\affiliation{Twinleaf LLC, 300 Deer Creek Drive, Plainsboro, NJ 08536, USA}

\author{A. Kryemadhi}
\affiliation{Department of Computing, Math \& Physics, Messiah University, Mechanicsburg PA 17055, USA}

\author{N. Kukowski}
\affiliation{Friedrich Schiller University Jena, Institute of Geosciences, Burgweg 11, D-07749 Jena, Germany}

\author{G. {\L}ukasiewicz}
\affiliation{Institute of Physics, Jagiellonian University in Krakow, prof. Stanis{\l}awa {\L}ojasiewicza 11, 30-348, Krak\'ow, Poland}
\affiliation{Jagiellonian University, Doctoral School of Exact and Natural Sciences, Łojasiewicza 11, 30-348, Krak\'ow, Poland}

\author{H. Masia-Roig}
\affiliation{Helmholtz Institute Mainz, 55099 Mainz, Germany}
\affiliation{GSI Helmholtzzentrum für Schwerionenforschung GmbH, 64291 Darmstadt, Germany}
\affiliation{Johannes Gutenberg-Universit{\"a}t Mainz, 55128 Mainz, Germany}

\author{M. Padniuk}
\affiliation{Institute of Physics, Jagiellonian University in Krakow, prof. Stanis{\l}awa {\L}ojasiewicza 11, 30-348, Krak\'ow, Poland}

\author{C. A. Palm}
\affiliation{Department of Physics, California State University -- East Bay, Hayward, California 94542-3084, USA}

\author{S. Y. Park}
\thanks{present address: Department of Physics, University of Colorado, Boulder, Colorado, 80309, USA}
\affiliation{Department of Physics and Astronomy, Oberlin College, Oberlin, OH 44074, USA}

\author{X. Peng}
\affiliation{State Key Laboratory of Advanced Optical Communication Systems and Networks, Department of Electronics, and Center for Quantum Information Technology, Peking University, Beijing 100871, China}

\author{M. Pospelov}
\affiliation{School of Physics and Astronomy, University of Minnesota, Minneapolis, MN 55455, USA}
\affiliation{William I. Fine Theoretical Physics Institute, School of Physics and Astronomy, University of
Minnesota, Minneapolis, MN 55455, USA}

\author{S. Pustelny}
\affiliation{Institute of Physics, Jagiellonian University in Krakow, prof. Stanis{\l}awa {\L}ojasiewicza 11, 30-348, Krak\'ow, Poland}

\author{Y. Rosenzweig}
\affiliation{Department of Physics, Ben-Gurion University of the Negev, Be’er Sheva 84105, Israel}

\author{O. M. Ruimi}
\affiliation{Racah Institute of Physics, Hebrew University of Jerusalem, Jerusalem 9190401, Israel}
\affiliation{Helmholtz Institute Mainz, 55099 Mainz, Germany}
\affiliation{GSI Helmholtzzentrum für Schwerionenforschung GmbH, 64291 Darmstadt, Germany}
\affiliation{Johannes Gutenberg-Universit{\"a}t Mainz, 55128 Mainz, Germany}

\author{P. C. Segura}
\thanks{present address: Department of Physics, Harvard University, Cambridge, Massachusetts, 02138, USA}
\affiliation{Department of Physics and Astronomy, Oberlin College, Oberlin, OH 44074, USA}

\author{T. Scholtes}
\affiliation{Leibniz Institute of Photonic Technology, Albert-Einstein-Straße 9, D-07745 Jena, Germany}

\author{Y. K. Semertzidis}
\affiliation{Center for Axion and Precision Physics Research, IBS, Daejeon 34051, Republic of Korea}
\affiliation{Department of Physics, KAIST, Daejeon 34141, Republic of Korea}

\author{Y. C. Shin}
\affiliation{Center for Axion and Precision Physics Research, IBS, Daejeon 34051, Republic of Korea}

\author{J. E. Stalnaker}
\affiliation{Department of Physics and Astronomy, Oberlin College, Oberlin, OH 44074, USA}

\author{D. Tandon}
\thanks{present address: Department of Physics, Stanford University, Stanford, California, 94305, USA}
\affiliation{Department of Physics and Astronomy, Oberlin College, Oberlin, OH 44074, USA}

\author{A.~Weis}
\affiliation{Physics Department, University of Fribourg, CH-1700 Fribourg, Switzerland}

\author{A. Wickenbrock}
\affiliation{Helmholtz Institute Mainz, 55099 Mainz, Germany}
\affiliation{GSI Helmholtzzentrum für Schwerionenforschung GmbH, 64291 Darmstadt, Germany}
\affiliation{Johannes Gutenberg-Universit{\"a}t Mainz, 55128 Mainz, Germany}

\author{T. Z. Wilson}
\thanks{present address: Department of Physics, University of Illinois Urbana-Champaign, Urbana, Illinois, 61801, USA}
\affiliation{Department of Physics, California State University -- East Bay, Hayward, California 94542-3084, USA}

\author{T. Wu}
\affiliation{State Key Laboratory of Advanced Optical Communication Systems and Networks, Department of Electronics, and Center for Quantum Information Technology, Peking University, Beijing 100871, China}

\author{J. Zhang}
\affiliation{State Key Laboratory of Advanced Optical Communication Systems and Networks, Department of Electronics, and Center for Quantum Information Technology, Peking University, Beijing 100871, China}

\author{Y. Zhao}
\affiliation{State Key Laboratory of Advanced Optical Communication Systems and Networks, Department of Electronics, and Center for Quantum Information Technology, Peking University, Beijing 100871, China}


\begin{abstract}
The history of astronomy has shown that advances in sensing methods open up new windows to the universe and often lead to unexpected discoveries.  Quantum sensor networks in combination with traditional astronomical observations are emerging as a novel modality for multi-messenger astronomy.  Here we develop a generic analysis framework that uses a data-driven approach to model the sensitivity of a quantum sensor network to astrophysical signals as a consequence of beyond-the-Standard Model (BSM) physics.   The analysis method evaluates correlations between sensors to search for BSM signals coincident with astrophysical triggers such as black hole mergers, supernovae, or fast radio bursts. Complementary to astroparticle approaches that search for particlelike signals (e.g. WIMPs), quantum sensors are sensitive to wavelike signals from exotic quantum fields.  This analysis method can be applied to networks of different types of quantum sensors, such as atomic clocks, matter-wave interferometers, and nuclear clocks, which can probe many types of interactions between BSM fields and standard model particles.

We use this analysis method to carry out the first direct search  utilizing a terrestrial network of precision quantum sensors for BSM fields emitted during a black hole merger.  Specifically we use the Global Network of Optical Magnetometers for Exotic physics (GNOME) to perform a search for exotic low-mass field (ELF) bursts generated in coincidence with a gravitational wave signal from a binary black hole merger (GW200311\_115853) detected by LIGO/Virgo on the 11th of March 2020.  The associated gravitational wave heralds the arrival of the ELF burst that interacts with the spins of fermions in the magnetometers. This enables GNOME to serve as a tool for multi-messenger astronomy. Our search found no significant events, and consequently we place the first lab-based limits on combinations of ELF production and coupling parameters.

\end{abstract}

\maketitle

\section{Introduction}
\subsection{Overview}
The Standard Model of particle physics is one of the great successes of modern science, providing a nearly perfect description of the particles that make up the stars, planets, and everyday objects around us.  However, astrophysical observations have shown that this ordinary matter makes up only a small fraction of our universe \cite{Planck2018,Kazin}.  These observations strongly hint at the existence of undiscovered  particles and fields beyond the Standard Model (BSM).   

The last few decades have seen the development of sensors utilizing the unprecedented control of ordinary matter through the tools of quantum mechanics \cite{Ye2024}.  Objects, ranging from single atoms to increasingly complex molecules \cite{gerlich2011quantum}, can be placed into specific quantum states.  This ability to tame the quantum nature of matter has opened up a vast toolbox enabling applications ranging from powerful quantum computers to a variety of quantum sensors with extreme precision.  By clever choice of atomic species and experimental design, precise measurements of accelerations, rotations \cite{Kimball2023PRA}, and electric and magnetic fields  are possible.  These precise quantum sensors utilizing ordinary atoms can in turn be used to search for small effects from interactions with undiscovered BSM particles and quantum fields \cite{safronova2018search}. Quantum sensors are typically tabletop in size and can be deployed in networks across the world.  Astrophysical signals show up as correlations across global networks allowing for the rejection of spurious backgrounds.   Having a diversity of sensors can be an advantage, providing a unique fingerprint for signals from BSM physics.

Our method can search for BSM signals coincident with triggers from astronomical observations such as gravitational wave signals from black hole mergers or optical signals from fast radio bursts and even supernovae \cite{o2013interpreting}.  These are some of the highest energy events in the universe and provide a unique gateway for the production of BSM particles and fields in environments not accessible in Earth laboratories.

\subsection{Application to ELFs}
Many extensions of the Standard Model posit the existence of new scalar, pseudoscalar, vector, or pseudovector fields with light ($m \ll 1\,\rm{keV /c^2}$) quanta \cite{antypas2022new}, such as axions and axion-like particles (ALPs) \cite{PRESKILL1983127,ABBOTT1983133,DINE1983137,Arv10} and dark/hidden photons \cite{ackerman2009dark,goodsell2009naturally,nelson2011dark}. In this work, we apply our general method to search for light, ultra-relativistic pseudoscalar fields triggered by an astrophysical event using the Global Network of Optical Magnetometers for Exotic physics searches (GNOME). Following Ref.\,\cite{Dailey2021}, we refer to these fields as exotic low-mass fields (ELFs). We define ELFs, in general, to be any BSM low-mass ($m \ll 1\,\rm{keV /c^2}$) bosonic field (spin-0 or spin-1). We do not assume ELFs constitute a significant fraction of dark matter and thus ELF parameters are not directly constrained by dark matter searches. Here we focus on a specific class of ELFs, namely spin-0 axion or axion-like fields that couple to fermion spins, yet the analysis method described here can be extended in straightforward ways to other classes of ELFs.\footnote{Recently (during the publication process of this paper), a closely related search for ELFs was carried out using the rubidium clocks onboard satellites of the GPS \cite{sen2024multi}. The search described in Ref.\,\cite{sen2024multi} targeted a different interaction of ELFs with atoms than that probed in the present study: transient variations in fundamental constants. Analyzing GPS clock data around the GW170817 neutron star merger, the authors of Ref.\,\cite{sen2024multi} identify an intriguing post-trigger excess in clock noise, although elevated solar electron flux during the same period complicates interpretation. This prevented them from being able to set constraints on ELF production mechanisms or interactions with atoms.} ELFs are postulated to be generated during large energy astrophysical events such as supernovae~\cite{Raf99}, binary black hole (BBH) or binary neutron star mergers \cite{baumann2019probing}, and fast radio bursts (FRB)~\cite{Tka15}. The detection of associated gravitational waves (GW) or electromagnetic (EM) signals herald the ELF signal arrival on Earth, and also indicates the GW/EM and ELF source location in the sky. By searching for signatures of the emissions from known sources shortly after a GW or EM trigger, we set limits on possible ELF production and coupling parameters associated with the observed astrophysical event. 

The production mechanism of ELFs is an open question~\cite{arvanitaki2017black,baryakhtar2017black,baumann2019probing,baryakhtar2021black, takahashi2024self,aurrekoetxea2024self, eby2022probing}. Black holes may be surrounded by clouds of exotic bosons, with up to 10\% of the black hole mass in the clouds~\cite{arvanitaki2015discovering}. Theories of scalar-tensor gravity~\cite{Fujii:2003pa} describe cases of black holes and neutron stars being immersed in scalar fields. Such scenarios suggest that modes of these scalar fields  may be excited during BBH or binary neutron star mergers~\cite{franciolini2019effective}. To set limits on ELF properties, we proceed with the following generic assumptions: (1) Some high-energy astrophysical event emits a spherical wave packet of the ELF $\phi(r)$ from a known location at distance $R$ from Earth. (2) The emitted wave packet carries away total energy $\Delta E$, where $\Delta E/\varepsilon_0 \gg 1$ with $\varepsilon_0$ being the energy of an ELF quanta. Given the assumptions above, the ELF will have high mode occupation number and hence they are treated as classical phase-coherent waves. We place the first lab-based constraints which can be used to limit possible ELF models as well as their production mechanisms.

\subsection{Outline}

We present a generic analysis framework to search for correlated ELF signals in a quantum sensor network.  We take a data-driven approach, using the observed background noise of the sensors to model the network sensitivity to astrophysical signals from BSM physics. We apply this approach to the specific case of searching for ELFs coincident with a GW detection with the GNOME network.

This manuscript is organized as follows. In Sec.~\ref{sec:model}, we recall the characteristics of an ELF pulse, its propagation, and its coupling to atomic magnetometers. In Sec.~\ref{sec:network}, we describe the GNOME network. In Sec.~\ref{sec:pipeline}, we present our two-stage analysis pipeline. First, we look for large excess power in the network at particular times and frequencies. We then perform likelihood-ratio tests using the characteristics of the individual stations to determine the significance of possible events compared to time-shifted data, where we expect no signals. We perform the same test on injected signals to calibrate the network response to a known ELF signal. Finally, in Sec.~\ref{sec:results}, we present results for one search target: the BBH merger GW200311\_115853 detected by LIGO/Virgo \cite{GCN_BBH}. 

\begin{figure}[h!]
    \centering
    \includegraphics[width =\linewidth]{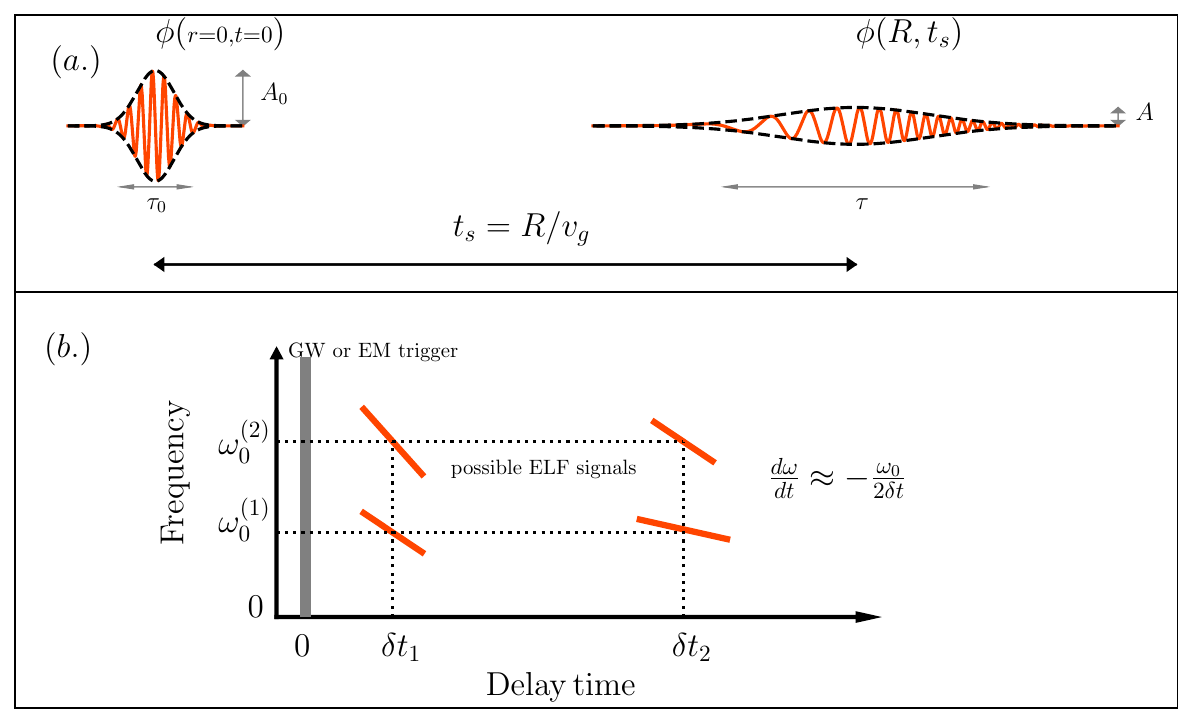}
    \caption{ELF propagation and signal characteristics. (a) An ELF wave packet, $\phi(r,t)$, emitted from a source at position $r=0$ and time $t=0$ with amplitude $A_0$ and pulse duration $\tau_0$ disperses as it propagates to the sensor through the distance $R$ and arrival time $t_s$ with group velocity $v_g$. At the sensor, the ELF wave packet has amplitude $A$ and pulse duration $\tau$. (b) Examples of ELF signal spectrograms with two possible central frequencies, $\omega_0^{(1)}$ and $\omega_0^{(2)}$, and two possible delay times, $\delta t_1$ and $\delta t_2$, with respect to the arrival of the GW/EM trigger.  Dispersion leads to a frequency chirp $d\omega / dt$ of the signal depending on the arrival time $\delta t$ and the central frequency $\omega_0$ [(Eq.\,\eqref{Eq:Slope}].}
    \label{fig:ELFcartoon}
\end{figure}

\section{Models for ELF propagation and Detection} \label{sec:model}

\subsection{Propagation} \label{sub: prop}

We start with a cartoon picture for the ELF-signal propagation as shown in Fig.\,\ref{fig:ELFcartoon}. An ELF wave packet is released at the time of an astrophysical event, which emits a GW or EM pulse. As it travels towards the sensors on Earth, the wave packet disperses while conserving its total energy.  This leads to an observed frequency chirp at the detector as discussed in more detail below.  

Formally, the scalar ELF $\phi$ can be described by the Klein-Gordon equation \cite{jacksonkimball2022}.  In the case of spherical symmetry, this has 1D solutions of the form $\phi(r,t) \propto \exp( i k r  \pm i \omega  t )/r$, where the frequency $\omega(k) = \left[ (ck)^2  + (mc^2/\hbar)^2\right]^{1/2}$  follows the relativistic dispersion relation for the wavenumber $k$, the momentum $\hbar k$, and the mass of the quanta of the scalar field $m$. 

Following the derivation in Ref.~\cite{Dailey2021}, we consider an ELF pulse having total energy $\Delta E$, and assume it has a Gaussian envelope with central ELF frequency $\omega_0$ and temporal length $\tau_0$. After propagation through space, the ELF pulse arrives at Earth with a delay $\delta t$ after the GW/EM pulse. The waveform at the detector will take the form:

\begin{align}
\phi(t) \approx \frac{A_0}{R}\, \sqrt{\frac{\tau_0}{\tau}}\exp\prn{-\frac{(t-t_s)^2}{2 \tau^2}} \cos\left[ \omega_0 (t-t_s) - \frac{\omega_0}{4\delta t}(t-t_s)^2 \right]\,,
\label{Eq:DispersionModel}
\end{align}
where $t_s$ is the time of arrival of the ELF pulse, $\tau$ is the temporal length at the detector, $R$ is the distance from the source, and the amplitude $A_0$ is given by:

\begin{equation}
    A_0  \approx \frac{1}{\pi^{1/4}}\prn{\frac{1}{\omega_0}\sqrt{\frac{c\Delta E}{2\pi \tau_0}}}\, .
    \label{App:Eq:AmplitudeGaussian}
\end{equation}
The second term in the cosine argument of Eq.\,\eqref{Eq:DispersionModel} comes from dispersion of the frequency components of the pulse, leading to a chirped signal.

For an ultrarelativistic field the characteristic energy of the ELF quanta $\epsilon_0$ is dominated by the kinetic energy, $\varepsilon_0 = \hbar \omega_0\approx c\hbar k_0$ (see Ref.~\cite{Dailey2021} for details).  By expanding the dispersion $\omega(k)$ around $k_0$ we identify the group velocity 
\begin{equation}
\frac{v_g}{c} = \frac{1}{c}\left.\frac{\partial\omega}{\partial k}\right|_{k=k_0}\!=\frac{ck_0}{\omega_0}\approx\, 1-  \frac{1}{2}  \left( \frac{m c^2}{\varepsilon_0} \right)^2  \, ,   \label{App:Eq:group}
\end{equation}
which leads to a delay of the arrival of the ELF pulse on Earth relative to the GW or EM trigger given by 
\begin{equation}
\delta t = t_s-\frac{R}{c}=R\left(\frac{1}{v_g} - \frac{1}{c} \right) \approx \left( \frac{m c^2}{\varepsilon_0} \right)^2  \frac{R}{2c}. 
\label{App:Eq:Lag}
\end{equation}
Thus a given delay time and central frequency can be associated with a particular ELF mass. The spread in group velocities is
\begin{equation}
\frac{\Delta v_g}{c} = \left( \frac{m c^2}{\varepsilon_0} \right)^2  \frac{ \Delta \varepsilon} {\varepsilon_0},
\label{App:Eq:deltaGroup}
\end{equation}
where the energy spread from the source $\Delta\varepsilon=h \Delta \omega = \hbar / \tau_0$ leads to a frequency spread $\Delta\omega$. Due to the dispersion, the duration of the pulse at the sensors on Earth is given by

\begin{equation}
\tau(t_s) = \sqrt{\tau_0^2+\prn{\frac{\Delta v_g t_s}{v_g}}^2}\, = \sqrt{\tau_0^2+\prn{\frac{2\Delta \varepsilon \delta t}{\varepsilon_0}}^2}\, = \sqrt{\tau_0^2+\prn{\frac{2 \delta t}{\omega_0 \tau_0}}^2}\,.
\label{Eq:taufunction}
\end{equation}
In the limit when dispersion has significantly lengthened the pulse, $\tau\gg\tau_0$, Eq.\,\eqref{Eq:taufunction} is approximated by
\begin{equation}
\tau \approx \frac{2 \delta t}{\omega_0 \tau_0},
\label{Eq:duration}
\end{equation}
which we call the ``chirp approximation". This leads to a chirp of the frequency at the sensors on Earth, given by the derivative of the instantaneous frequency in Eq.\,\eqref{Eq:DispersionModel}:
\begin{align}
\frac{d\omega(t)}{dt} & = -\frac{\omega_0}{2\delta t},    \label{Eq:Slope}     
\end{align}
that depends only on the central frequency $\omega_0$ and delay time $\delta t$.

\subsection{ELF detection with magnetometers}\label{sub: detect}
We consider both linear and quadratic interactions of the ELF $\phi$ with our sensors: 
 \begin{align}
\mathcal{L}^{(l)} &= f_l^{-1} J^\mu \partial_\mu \phi \, \label{Eq:Lag1},  \\
 \mathcal{L}^{(q)} &= f_q^{-2} J^\mu \partial_\mu \phi^2\, , \label{Eq:Lag2}
\end{align}
where $J^\mu = \bar{\psi} \gamma^\mu \gamma_5 \psi$ is the axial-vector current for Standard Model fermions and $f_l$ and $f_q$ are the characteristic energy scales of the linear and quadratic interactions, respectively.  Transforming to a Hamiltonian representation, we arrive at the effective spin-dependent Hamiltonians 
\begin{equation}
\begin{aligned}
H^{(l)} & \approx -
\frac{2 (\hbar c)^{3/2} }{f_l} \mathbf{S} \cdot \mathbf{\nabla} \phi  
,  \\
H^{(q)} & \approx -
\frac{2 (\hbar c)^{2}}{f_q^2} \mathbf{S} \cdot \mathbf{\nabla} \phi^2, 
\label{Eq:magnetometer-nonrel-Hamiltonians}
\end{aligned} 
\end{equation}
with $\bf{S}$ being the atomic or nuclear spin \cite{Afach2023}. The ELF can be treated as a so-called ``pseudomagnetic field'' since it causes energy shifts of Zeeman sublevels in a manner similar to the effect of a real magnetic field.\footnote{In the present work we focus on the direct coupling of the ELF field $\phi$ with fermion spins \cite{cong2024spin}, but interactions with other standard model particles and fields are possible, see, e.g., Refs.\,\cite{safronova2018search,jacksonkimball2022}. While GNOME magnetometers \cite{afach2018characterization} and especially noble-gas comagnetometers \cite{PhysRevResearch.6.013339_comag} have state-of-the-art sensitivity to $\phi$-spin couplings \cite{jackson2023probing}, other quantum sensors such as atomic clocks, cavities, and atom interferometers are better suited for detection of other types of $\phi$-couplings \cite{safronova2018search,jacksonkimball2022}. One example is the search for ELF-induced transient variation of fundamental constants using the GPS atomic clock network described in Ref.\,\cite{sen2024multi}. Another relevant example is the $\phi$-photon coupling \cite{sikivie1983experimental}, which can lead to generation of a {\emph{real}} oscillating magnetic field as opposed to a pseudomagnetic field. In this case the magnetometer need not be spin-based, and other sensors such as superconducting quantum interference devices (SQUIDs) \cite{brouwer2022proposal} or induction-coil magnetometers \cite{sulai2023hunt} can be used. Of course, the dependence of the measured signal on $\phi$ will depend on the details of the $\phi$-coupling and the sensor characteristics: for example, as discussed in Refs.\,\cite{Dailey2021,arakawa2025multimessenger}, a scalar $\phi$-coupling that generates a signal detectable with atomic clocks does not have the directionality characteristic of the $\phi$-spin coupling considered here.}  To explicitly make the analogy between the Zeeman Hamiltonian ($H_Z = -\boldsymbol{\mu} \cdot \boldsymbol{B}$, where $\boldsymbol{\mu} = \mu \boldsymbol{S}$ is the magnetic moment and $\boldsymbol{B}$ is a magnetic field) and the Hamiltonians describing ELF-spin interactions, we introduce the corresponding pseudomagnetic fields:
\begin{equation}
\begin{aligned}
\Bp^{(l)} = \frac{2 (\hbar c)^{3/2}}{\mu_B f_l} \nabla \phi ,\\
\Bp^{(q)} = \frac{2 (\hbar c)^{2}}{\mu_B {f_q}^2} \nabla \phi^2,
\end{aligned}
\end{equation}
where the Bohr magneton $\mu_B$ is introduced as the unit of magnetic moment. To lowest order in $1/R$, one can show that the gradients of the waveform [Eq.\,\eqref{Eq:DispersionModel}] at the detectors are: 
\begin{equation}
    \nabla \phi(R,t_s) \approx \frac{\hat{R}}{R} \sqrt{\frac{\Delta E}{2 \pi^{3/2} c  \tau}}
    \label{Eq:grad_phi}
\end{equation}
 and

\begin{equation}
    \nabla \phi^2(R,t_s) \approx \frac{\hat{R}}{R^2}\frac{\Delta E}{2 \pi^{3/2} \omega_0 \tau},
    \label{Eq:grad_phisquared}
\end{equation}
where $\hat{R}$ is the unit vector pointing from the detector to the event location.

GNOME is designed to detect exotic spin interactions via the correspondence with the Zeeman interaction. For a given energy scale $f^{l(q)}$, the above considerations imply that an ELF incident on the GNOME network will manifest as a transient pseudomagnetic field $\Bp$.   If the network has sufficient sensitivity, this will result in coincident detection of a pseudomagnetic field with the characteristic network pattern given by the orientations of different sensors.  Alternatively, given a null detection, we can set limits on combinations of $f^{l(q)}$ and the unknown initial energy release $\Delta E$ via

\begin{equation}
    f_l\times \left(\sqrt{\frac{1}{\Delta E}}\right) \gtrsim \frac{\hbar^{3/2} \, c}{\mu_B \, \Bp^* R} \sqrt{\frac{2}{\pi^{3/2} \, \tau}},
\label{Eq:fl_lim}
\end{equation}

\begin{equation}
    f_q\times  \left(\sqrt{\frac{1}{\Delta E}}\right) \gtrsim \frac{\hbar c}{R}\,\sqrt{\frac{1}{ \pi^{3/2} \mu_B \Bp^* \,\omega_0 \, \tau}},
\label{Eq:fq_lim}
\end{equation}
where the pulse duration at the detector $\tau$, given by Eq.\,\eqref{Eq:taufunction}, is evaluated at the time of arrival of the ELF pulse $t_s$ for a given $\tau_0$. The values of $\Delta E$ and $\tau_0$ can vary widely depending on the production theory \cite{arvanitaki2015discovering,arvanitaki2017black,baryakhtar2017black,baumann2019probing,baryakhtar2021black}. It is important to note that the observed frequency of the quadratic interaction corresponds to twice the central ELF frequency since the gradient of the field squared couples to the spin.

In the limit when significant dispersion has lengthened the ELF pulse, $\tau$ can be evaluated with the chirp approximation [Eq.\,\eqref{Eq:duration}] and the exclusion limits scale with the initial pulse duration by $\tau_0^{1/2}$. Without appreciable dispersion or lengthening of the pulse, $\tau\approx \tau_0$, the exclusion limits scale roughly as $\tau_0^{-1/2}$.

\section{GNOME Network Characteristics}\label{sec:network}
 GNOME comprises a global network of magnetically shielded alkali-atomic magnetometers \cite{pustelny2013global,afach2018characterization,Afach2021,Afach2023}. These are located in more than a dozen stations distributed over four continents.  The sensors measure the projection of the ambient magnetic field on one sensitive axis via the variation of the Larmor spin precession of optically polarized alkali atoms in a vapor cell. Measurements are time-stamped using the Global Position System allowing for the use of coincident detection techniques to suppress unwanted background noises \cite{Afach2021}. In addition, a multi-layer mu-metal magnetic shield, present in each station, suppresses background magnetic fields by a factor of $\gtrsim10^6$. Because most GNOME magnetometers use atoms whose nuclei have one unpaired proton spin, they are primarily sensitive to proton-spin couplings with the ELF. The expected ELF signal amplitudes measured by the sensors will be proportional to the relative contribution of the proton spin to the nuclear spin \cite{Afach2021}. ELFs can also couple to electron spins but the signal is attenuated by the electrons in the magnetic shields. To first order it was expected this attenuation is the same as the magnetic shielding factor \cite{kimball2016magnetic} but the orbital angular momentum of the electrons in the shields can lessen this effect \cite{futurePaper}. The coupling to electron spins is ignored in this manuscript.

In Table \ref{table: MagInfo}, we present some characteristics of the sensors in the network. More detailed descriptions of the sensors and the network are given in Ref.~\cite{afach2018characterization}. 
\begin{table}[h!]
\centering
\caption{The station name, location in longitude and latitude, orientation of the sensitive axis, and atomic species used in GNOME sensors. The rightmost column lists the estimated ratio between the effective proton spin polarization \cite{Kim15} and the Land\'e $g$-factor for the magnetometer, $\sigma_p/g$, which depends on the atomic species and the magnetometry scheme as described in Appendix~B of Ref.~\cite{Afach2021_domainwall}. * marking denotes the stations that were active with quality data during the GW200311\_115853 event. (\textsuperscript{\textdagger}Moved from Fribourg to Moxa in 2018.)}
\begin{tabular}{lrrrrrcc}
\hline
\hline
          &                  \multicolumn{3}{c}{Location}                 &           \multicolumn{2}{c}{Orientation}           & \multicolumn{1}{c}{} \\
Station   &  \multicolumn{1}{l}{$\qquad$Longitude} & & \multicolumn{1}{l}{Latitude} &   \multicolumn{1}{c}{$\qquad$Az} & \multicolumn{1}{c}{Alt}  & Species & \multicolumn{1}{c}{$\sigma/g_f$}  \\ 
\hline 
\rule{0ex}{3.0ex}Beersheba   & 34.8043\textrm{\textdegree~E} & &31.2612\textrm{\textdegree~N} & 180\textrm{\textdegree} &   0\textrm{\textdegree} & $^3$He/$^{87}$Rb/$^{39}$K & see footnote \footnote{The Beersheba station employs a $^3$He/K/Rb SERF comagnetometer similar to Ref.\,\cite{klinger2023optimization}. The conversion between magnetic field and pseudomagnetic field for such a comagetometer, described in Ref.\,\cite{PhysRevResearch.6.013339_comag}, is frequency dependent and therefore not described by a single number $\sigma/g_f$.} \\
\rule{0ex}{3.0ex}Beijing   & 116.1868\textrm{\textdegree~E} & &40.2457\textrm{\textdegree~N} & 251\textrm{\textdegree} &   0\textrm{\textdegree} & $^{133}$Cs&$-0.39^{+0.19}_{-0.00}$ \\
\rule{0ex}{3.0ex}Berkeley  & 122.2570\textrm{\textdegree~W} & &37.8723\textrm{\textdegree~N} &    0\textrm{\textdegree} & 90\textrm{\textdegree} & $^{133}$Cs&$-0.39^{+0.19}_{-0.00}$\\
\rule{0ex}{3.0ex}Canberra &  149.1185\textrm{\textdegree~E} & & 35.2745\textrm{\textdegree~S} &   0\textrm{\textdegree} & 90\textrm{\textdegree} & $^{87}$Rb&$\ \ 0.70^{+0.00}_{-0.15}$\\  
\rule{0ex}{3.0ex}Daejeon   & 127.3987\textrm{\textdegree~E} & &36.3909\textrm{\textdegree~N} &    0\textrm{\textdegree} & 90\textrm{\textdegree} & $^{133}$Cs&$-0.39^{+0.19}_{-0.00}$\\ 
\rule{0ex}{3.0ex}Hayward*   & 122.0539\textrm{\textdegree~W} & &37.6564\textrm{\textdegree~N} &    0\textrm{\textdegree} & 90\textrm{\textdegree} &$^{87}$Rb &$\ \ 0.70^{+0.00}_{-0.15}$\\ 
\rule{0ex}{3.0ex}Hefei     & 117.2526\textrm{\textdegree~E} & &31.8429\textrm{\textdegree~N} &  90\textrm{\textdegree} &   0\textrm{\textdegree} &  $^{85}$Rb/$^{87}$Rb & $-0.38^{+0.05}_{-0.00}$\\ 
\rule{0ex}{3.0ex}Krakow    &  19.9048\textrm{\textdegree~E} & &50.0289\textrm{\textdegree~N} &  45\textrm{\textdegree} &   0\textrm{\textdegree} & $^{87}$Rb&$\ \ 0.50^{+0.00}_{-0.11}$\\ 
\rule{0ex}{3.0ex}Lewisburg*   &  76.8825\textrm{\textdegree~W} & &40.9557\textrm{\textdegree~N} &    0\textrm{\textdegree} & 90\textrm{\textdegree} &$^{87}$Rb &$\ \ 0.70^{+0.00}_{-0.15}$\\  
\rule{0ex}{3.0ex}Los Angeles*     &  118.4407\textrm{\textdegree~W} & & 34.0705\textrm{\textdegree~N} &    270\textrm{\textdegree} & 0\textrm{\textdegree} & $^{85}$Rb &$\ \ 0.50^{+0.00}_{-0.07}$\\  
\rule{0ex}{3.0ex}Mainz     &   8.2354\textrm{\textdegree~E} & &49.9915\textrm{\textdegree~N} &    0\textrm{\textdegree} & -90\textrm{\textdegree} & $^{87}$Rb &$\ \ 0.50^{+0.00}_{-0.11}$\\
\rule{0ex}{3.0ex}Moxa*\textsuperscript{\textdagger}    &  11.6147\textrm{\textdegree~E} & &50.6450\textrm{\textdegree~N} &    270\textrm{\textdegree} & 0\textrm{\textdegree} &$^{133}$Cs & $-0.39^{+0.19}_{-0.00}$\\  
\rule{0ex}{3.0ex}Oberlin* &  81.7796\textrm{\textdegree~W} & &41.2950\textrm{\textdegree~N} &    276\textrm{\textdegree} & 0\textrm{\textdegree} & $^{39}$K &$-0.49^{+0.18}_{-0.00}$\\

\hline
\hline

\end{tabular}
\label{table: MagInfo}
\end{table}
For this work, each sensor has bandwidth of approximately 100\,Hz. Magnetic field sensitivities for the sensors in each station vary from around 0.01 to 1 $\mathrm{pT/\sqrt{Hz}}$ over this band. Example characterizations are given in \cite{afach2018characterization}. The response and sensitivity of each sensor is determined by calibration signals at multiple frequencies. Magnetic field measurements are recorded at a sampling rate of 512 samples per second. In addition, a 1 sample per second data glitch / data quality indicator is also written to file.  This boolean variable is derived from auxiliary sensors which monitor temperature, accelerations, magnetic field outside the shielding, rotation and other user-defined variables such as laser intensity, frequency, etc. If any of those quantities vary outside an allowed range within a given second, the data-quality flag -- which is normally true -- is switched to false. In our analysis, we disregard all the data having data quality flag false.

While an incident ELF yields a common pseudomagnetic field  $\Bp$, each particular sensor, denoted throughout with a superscript $i$, will sense an effective magnetic field $B^i$, which is proportional to the projection of $\Bp$ on the sensor's sensitive axis $\hat{m}^i \cdot \hat{R}$ and the ratio between the proton-spin coupling and the Land\'e $g$-factor $g_F$ ($\sigma^i/g^i_F$) \cite{Afach2021}. Thus a field $\Bp$ incident on the network results in data values $d_i(t)$ from a sensor $i$ of the form 
\begin{equation}
    d^i(t) = B^i(t) + n^i(t),
    \label{eq:sigplusnoise}
\end{equation} where 
\begin{equation}
B^i(t) = \frac{\sigma^{i} \hat{m}^i(t)\cdot \hat{R}}{g^i_{F}} \Bp(t)
\label{Eq:Bi to Bp}
\end{equation}
is the contribution of the signal from the ELF, and $n_i(t)$ represents the environmental and sensor noise. We have assumed above that the detector bandwidths are much greater than the frequency spread of the signal. Astrophysical observations provide the distance $R$ of the event as well as the direction $\hat{R}$ given by right ascension and declination coordinates.   

\begin{figure}[h!]
    \centering
    \includegraphics[width = \linewidth]{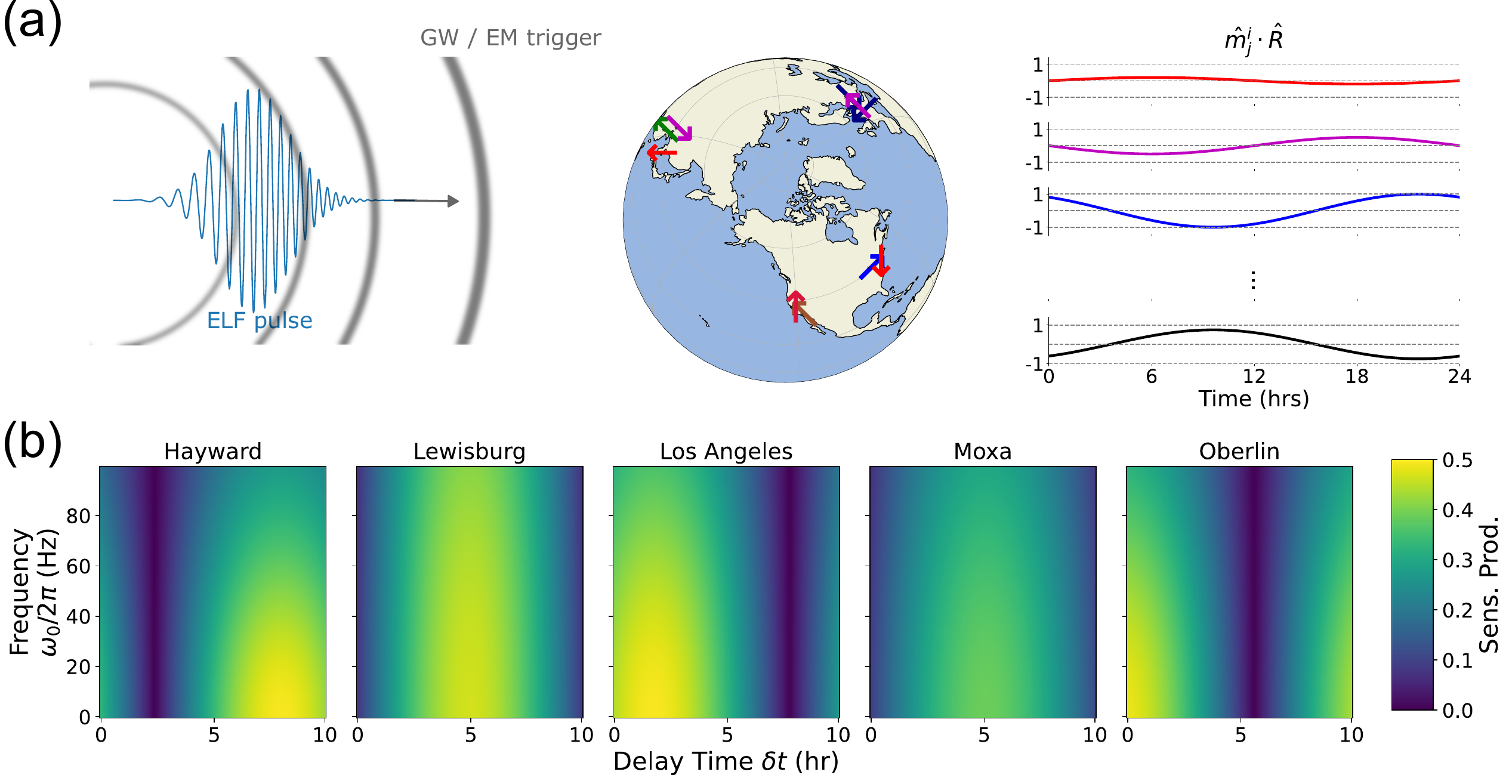}
    \caption{(a) Cartoon picture of the GNOME network with arrows representing GNOME detectors and their sensitive axes.  The arrival of a GW/EM trigger heralds the ELF pulse. The projection of the ELF on detector axes is modulated by Earth's rotation. (b)  Sensitivity product for each station which is the product of the directional sensitivity $\hat{m}^i_j\cdot \hat{R}$, frequency response $\beta^i_k$, and ratio of proton-spin coupling and the Lande g-factor $\sigma^i / g^i_F$ [Eq.\,\eqref{Eq:Bj to Bp, more indices}] across the foreground window for the BBH merger search target considered. We update the directional sensitivity at one minute intervals and the frequency response at 1\,Hz intervals.}
    \label{fig:dir_sen}
\end{figure}

 Because we are searching for an ELF signal with a mass-dependent central frequency and delay time, our analysis uses spectrograms formed by taking discrete Fourier transforms of subsets of the time series $d^i(t)$.  The output of the spectrogram is divided into ``tiles'' of the time-frequency plane.  We reserve the subscript $j$ for a tile's central time coordinate and the subscript $k$ for the central frequency coordinate.  The tile amplitudes are thus related to the time-domain signals via  
\begin{equation}
    \tilde{d}^i_{jk} = \tilde{B}^i_{jk} + \tilde{n}^i_{jk},
    \label{Eq:DFT}
\end{equation}
where $\tilde{B}^i_{jk}$ and $\tilde{n}^i_{jk}$ are the Fourier transform amplitudes at frequency index $k$ for a time subset denoted by $j$, and include the effects of the finite bandwidths of the detectors.  Therefore, for each tile the form of an ELF signal at  sensor $i$ is given by:
\begin{equation}
\tilde{B}^i_{jk} = \frac{\sigma^{i} \hat{m}^i_j\cdot \hat{R}}{g^i_{F}} \beta^i_k \Bpk,
\label{Eq:Bj to Bp, more indices}
\end{equation}
which has the same form as the equivalent expression in the time domain [Eq. \eqref{Eq:Bi to Bp}]. $\Bpk$ represents the amplitude of $\Bp$ in the frequency band associated with index $k$ and time $j$. $\beta^i_k$ are transfer functions accounting for the response at frequency $k$ of sensor $i$ and  are measured by applying magnetic fields of known frequency and amplitude. Figure~\ref{fig:dir_sen} displays an example of the projection of the ELF gradient on the sensitive axes of the magnetometers. The color maps are the individual station's scaling between $\Bp$ and $B^i$ [Eq.\,\eqref{Eq:Bj to Bp, more indices}] over the time-frequency space that is searched in this paper. 

For our analysis, we use spectrogram tiles of area $\Delta f \times \Delta t = 1$. As we discuss in Sec.~\ref{sub:AdaptiveTiling}, the aspect ratio of these tiles is adapted to match the expected characteristics of a signal at a specific delay time and frequency.
 
\section{ELF Analysis Pipeline}\label{sec:pipeline}

Our analysis focuses on a search for a signal recorded with GNOME that is coincident with an observation of an astrophysical event and consistent with the signal-propagation model described in Sec.~\ref{sec:model}. Figure~\ref{fig:AnalysisOutline} illustrates how data from the network are sent through the analysis pipeline -- leading to either the detection of ELF candidates or the determination of upper limits. The pipeline steps are:

\begin{figure}[h!!]
    \centering
    \includegraphics[width = \linewidth]{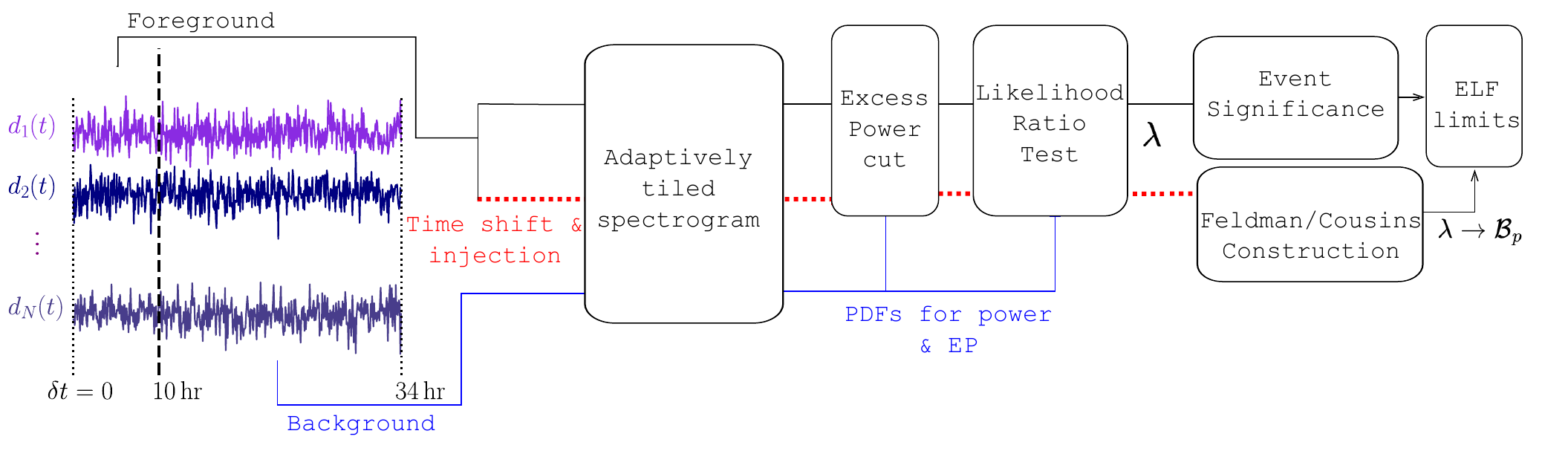}
    
    \caption{Flowchart illustrating the analysis pipeline. A search is triggered by some GW or EM observation. Time-series measurements from the network in the foreground and background windows are represented in the time-frequency plane by spectrograms with tile dimensions chosen to match the ELF chirp rate. The EP cut passes network-wide, relatively high power tiles to the next stage as candidate tiles. The LRT statistic $\lambda$ is determined for each candidate tile and is compared to a signal-free distribution of $\lambda$ to determine the significance. The test statistic is translated to a pseudomagnetic amplitude $\Bp$ with the Feldman-Cousins construction and ELF limits are calculated.}
    \label{fig:AnalysisOutline}
\end{figure}

 \begin{itemize}

    \item \textbf{Spectrogram}: Spectrograms of the time-series data from the network characterize the data in the time-frequency plane. The optimal temporal and frequency resolution is set by the arrival time of the GW/EM trigger and the expected-signal model. We tile the time-frequency plane adaptively to match the modeled signal characteristics as described in more detail in Sec.~\ref{sub:AdaptiveTiling}.

    \item \textbf{Excess power (EP) cut}:  For the first stage of the analysis, we perform a model agnostic EP cut \cite{Anderson2001} on the spectrograms to identify network-wide, relatively high power tiles labelled by their time-frequency coordinates. We apply a threshold on the joint EP probability of coincident tiles $\zeta^\epsilon$. This serves primarily to reduce the amount of data needing further analysis. We discuss this further in Sec.~\ref{sub:ExcessPower}.

    \item \textbf{Likelihood-ratio test (LRT)}:  Next, we perform a general LRT on the coincident tiles that passed the first stage \cite{James}. We fit the data to determine the most likely ELF pseudomagnetic field $\Bp$ and produce a test statistic $\lambda$ that quantifies the likelihood of an ELF signal + noise hypothesis against a noise-only hypothesis. We time-shift the data to effectively generate a large number of noise-only datasets which are used to determine the expected distributions of the test statistic. We discuss this further in Sec.~\ref{sub:Likelihood}.

    \item \textbf{Confidence-belt construction}:  To characterize the network sensitivity across the time-frequency space, we determine confidence belts using the Feldman-Cousins (FC) construction \cite{FeldCous}. We inject ELF signals with amplitude $\Bp$ into time-shifted data at coarse times and frequencies and perform the LRT to determine distributions of $\lambda$ in the presence of a signal. These confidence belts form a mapping between the test statistic $\lambda$ and the ELF pseudomagnetic field $\Bp$. We discuss this further in Sec.~\ref{sub:FeldCous}.

    \item \textbf{Event significance}: We determine the false positive probability of each candidate tile using the time-shifted distributions of the test statistic. We discuss this further in Sec.~\ref{sub:Significance}.

    \item \textbf{ELF detection or limits}:  If a tile with a significant test statistic is found the FC construction is used to translate this into a confidence belt on the ELF amplitude $\Bp$.  If no significant tiles are detected, we place an upper limit on the test statistic in each box by requiring the total false probability rate to be less than one over our dataset. We translate these to upper limits on the ELF amplitude via the FC construction. We then place limits on the combination of ELF couplings, the initial ELF-pulse duration, and energy release via the quantities $f_l\sqrt{1/ \Delta E}$ and $f_q\sqrt{1/ \Delta E}$ [Eqs.~\eqref{Eq:fl_lim}-\eqref{Eq:fq_lim}]. Section~\ref{sec:results} details our determination of ELF parameter upper limits.

\end{itemize}

The analysis combines methods from gravitational wave searches (EP, time-shifting) and standard statistical tools (LRT, FC construction, false positive probability) with the novel model-driven adaptively tiled spectrograms to search for ELFs with our quantum sensor network. We use data to make empirical distributions to determine the significance of candidate tiles. Note that the data analysis pipeline described here can be applied to networks of other quantum sensors searching for ELFs and also searches for ELFs composed of spin-1 particles (as opposed to the spin-0 particles assumed here) simply by appropriately modifying the function that describes the network response, e.g. Eq.~\eqref{Eq:Bj to Bp, more indices}.\footnote{Another significant difference between GNOME magnetometers and other quantum sensors is their bandwidth and sample rate. For example, state-of-the-art atomic clocks generally sample data at a much lower rate while optical cavities can sample data at a much higher rate. Even for the case of GNOME, another issue is sensor dead time. For this first analysis we chose to analyze a dataset where all sensors were contributing throughout.  Instead of a discrete Fourier transform, a future refinement to the algorithm could use a periodogram to analyze unevenly sampled data.}

\subsection{Spectrograms with adaptive tiling} 
\label{sub:AdaptiveTiling}
In time-frequency analysis methods, there is an inherent trade-off in localization in the time and frequency domains \cite{vetterli}. The tiling resolution for optimal signal-to-noise ratio is that which best matches the temporal and frequency spread of the signal and underlies methods such as wavelets and Q-transforms \cite{Chatterji_2004,Abbott_2020}. While we do not a priori know the duration of the ELF signal or its frequency spread, we do know that a signal arriving at a delay time and frequency coordinate $\{\delta t,\omega_0 \}$ will be chirped at a rate $d \omega /dt \approx -\omega_0/2 \delta t$ [Eq.\,\eqref{Eq:Slope}].  This suggests that the spectrogram tile dimensions should have the ratio 
\begin{equation}
    \frac{\Delta\omega}{\Delta t} \approx \frac{\omega_0}{2\delta t}\,.
    \label{Eq:tiledims}
\end{equation}  
Combining this constraint with the tile area given by the discrete Fourier transform, $\Delta \omega \Delta t = 2\pi$, gives unique tile dimensions over the delay time-frequency plane, which motivates our use of adaptive tiling.

\begin{figure}
    \centering
    \includegraphics[width = \linewidth]{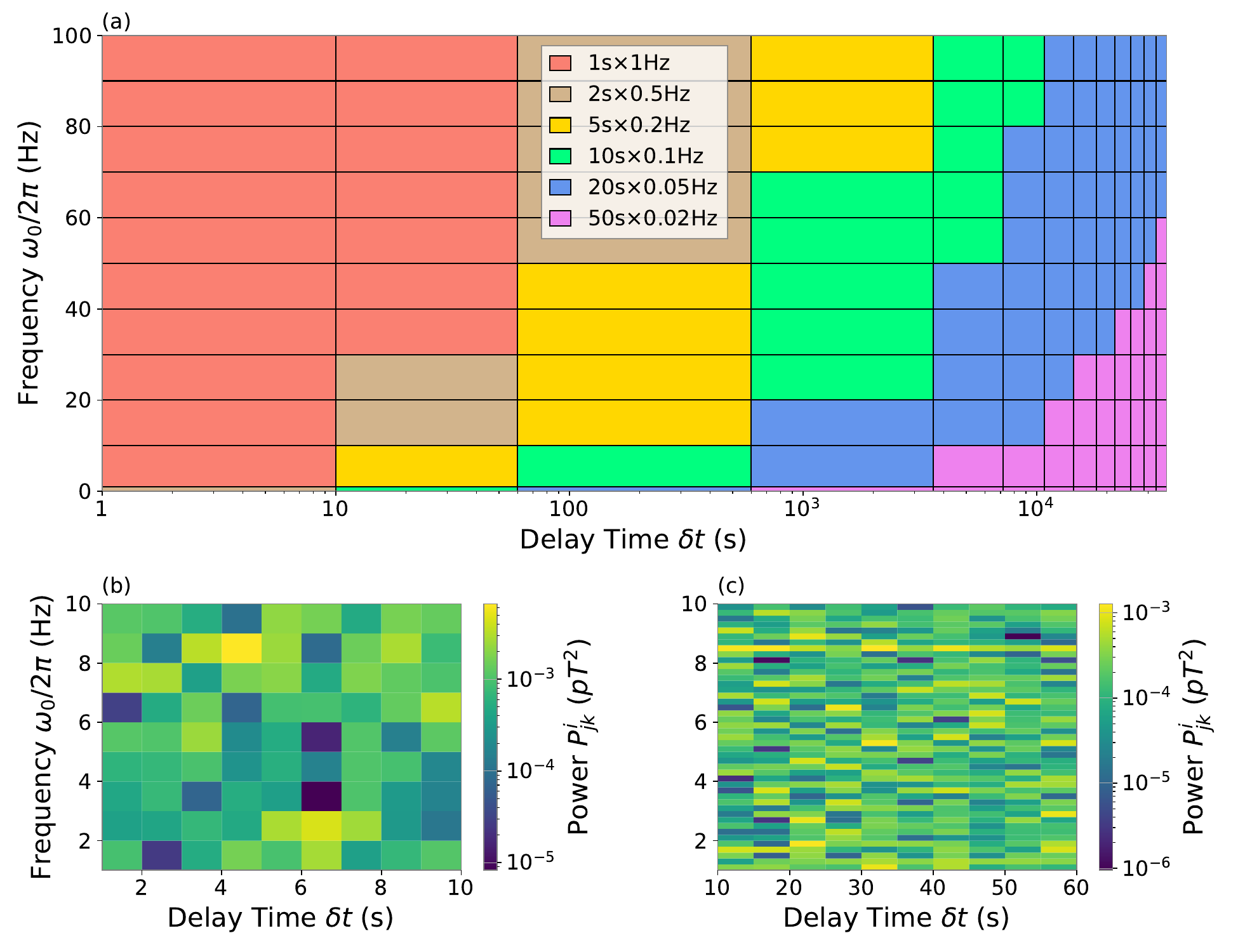}
    \caption{(a) Map of the adaptive tiling used to match the spectrogram tile dimensions to the chirp rate. The searched time-frequency space (foreground window), $\delta t \in [0,10]\,\text{hr} \times f_0 \in [0,100]$\,Hz, is partitioned into ``boxes" with boundaries marked by black lines. The box color indicates the tile dimension used that approximates the chirp rate given by Eq.\,\eqref{Eq:Slope}. (b) Spectrogram of Lewisburg data with tile dimension $1\,\text{s}\times 1\,\text{Hz}$ contained within the box spanning $f_0\in[1,10]\,\text{Hz}$ and $\delta t \in [1,10]\,\text{s}$. The relative color of each tile represents the measured power in the tile. (c) Spectrogram of Lewisburg data with tile dimension $5\,\text{s}\times 0.2\,\text{Hz}$ contained within the box adjacent to (b) spanning $f_0\in[1,10]\,\text{Hz}$ and $\delta t \in [10,60]\,\text{s}$.}
    \label{fig:Adaptive Tiling}
\end{figure}

We divide the foreground window into ``boxes" that span regions of delay time and frequency as shown in Fig.\,\ref{fig:Adaptive Tiling}.  For each box, we use tile dimensions, denoted by the box color, to approximate the ratio given by Eq.\,\eqref{Eq:tiledims}. Power spectrograms with the denoted tile dimensions are calculated in each box, giving the powers of individual tiles 
  \begin{equation}
      P^i_{jk}=\left|\tilde{d}^i_{jk}\right|^2, 
  \end{equation}
where $\tilde{d}$ are the amplitudes defined in Eq.\,\eqref{Eq:DFT} with $i$ denoting the sensor, $j$ the delay time, and $k$ the frequency. We do not consider the DC band as the ELF signal is an oscillating transient.

The discussion above applies when the ELF wave packet has dispersed as it propagates to Earth, i.e. $\tau \gg \tau_0$.  As discussed in Sec.~\ref{sub: prop}, this occurs when $2\delta t/\omega_0\tau_0\gg\tau_0$.  For the shortest delay times and highest frequencies this chirp approximation does not hold.  Here we do not apply any adaptive tiling and use the smallest tile dimension of $1\,\text{s}\times 1\,\text{Hz}$. 

\subsection{Excess power cut}
\label{sub:ExcessPower}

 The next step in our analysis is a cut based on EP. Excess power is a model-agnostic test statistic used to identify tiles that have power greater than the average noise \cite{Maggiore}.  Although less powerful than matched filtering, it provides an optimal search method for signals of unknown form \cite{Anderson2001}. In our analysis, a cut on the EP is performed primarily to reduce the amount of data to be processed by the model-dependent LRT. 
 
\begin{figure}[h!]
    \centering
    \includegraphics[width = \linewidth]{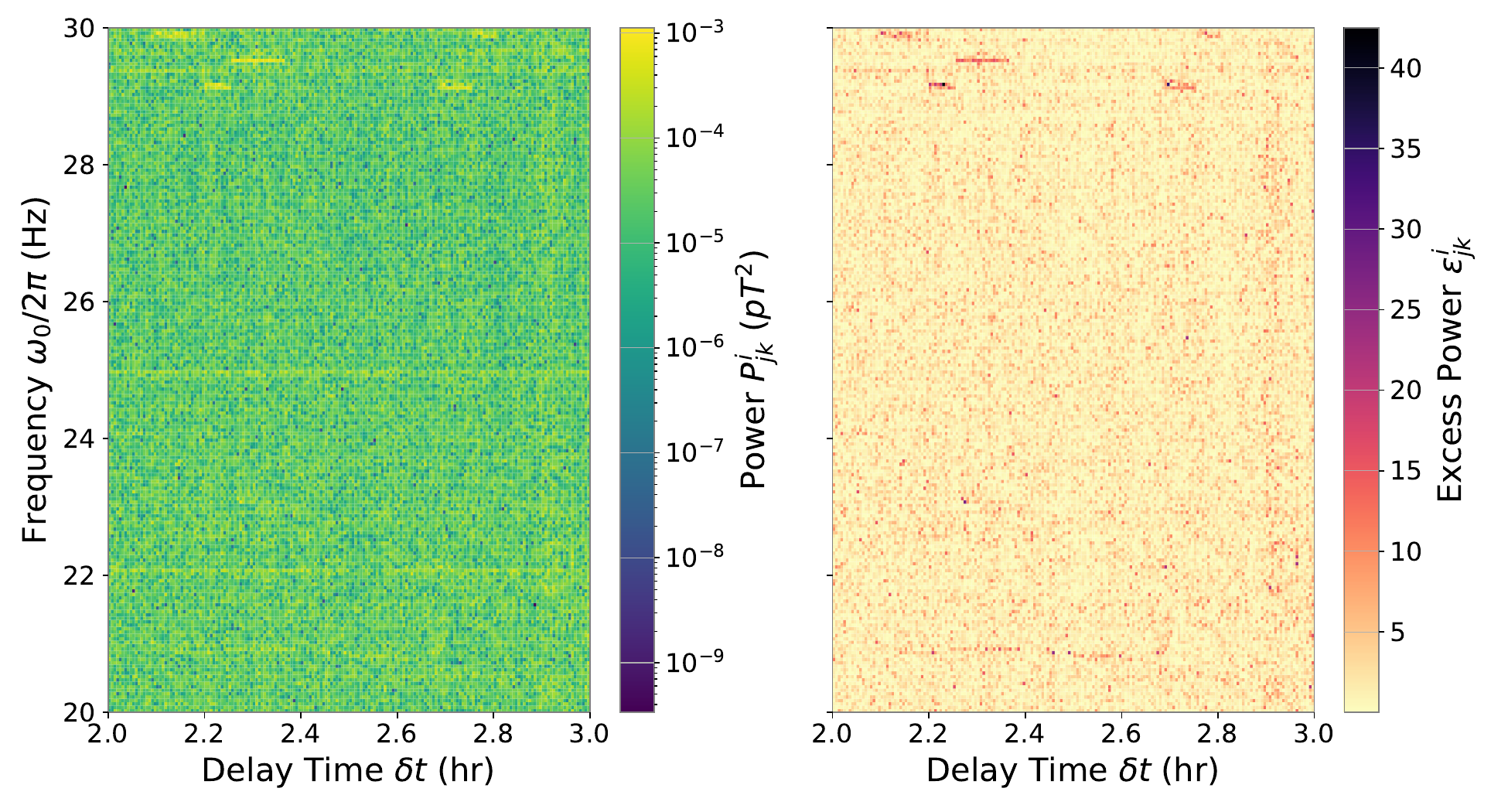}
    \caption{Whitening of spectrogram of $P^i_{jk}$ and corresponding EP statistic $\varepsilon^i_{jk}$ of Lewisburg data [Eq.\,\eqref{Eq: Excess Power}]. The relative color of each tile in each plot represents the measured power (on the left) and EP (on the right) in the tile. Note that the constant higher power bands at 22\,Hz and 25\,Hz are ``whitened'' to average EP values. The loud spikes in the 29\,Hz--30\,Hz range are mellowed due to their repetition while the quiet regions are especially suppressed due to the larger average noise. The features in the 29\,Hz--30\,Hz range indicate non-stationarity which is not suppressed by taking the EP.}
    \label{fig:P to EP}

\end{figure}

With our adaptive tiling, the tile dimension depends on which box is being analyzed.  Below, we focus on the analysis for a single tile dimension.  First, the unitless EP statistic $\varepsilon^i_{jk}$ for a tile of sensor $i$, at time $j$, and frequency $k$ is determined 
\begin{equation}
\varepsilon^i_{jk}=P^i_{jk} / \bar{P}^i_{k},
\label{Eq: Excess Power}
\end{equation}
where the measured power $P^i_{jk}$ is normalized by the average power $\bar{P}^i_k$ over the entire dataset for the respective frequency $k$. Figure~\ref{fig:P to EP} displays a contrast of the measured power and EP in a single time-frequency box for a single station.

In the presence of pure Gaussian noise, EP would be a chi-squared distributed statistic with two degrees of freedom. Our data, however, empirically deviates from this simple exponential, particularly in the tails of the probability density function (PDF), due to unique noise present in certain sensors. To avoid assumptions about Gaussianity when comparing stations, we construct empirical PDFs $f^i(\varepsilon)$ for each station.  The empirical PDFs are formed using data from a background window consisting of the following 24 hours and are frequency-independent. 

\begin{figure}[h!]
    \centering
    \includegraphics[width = \linewidth]{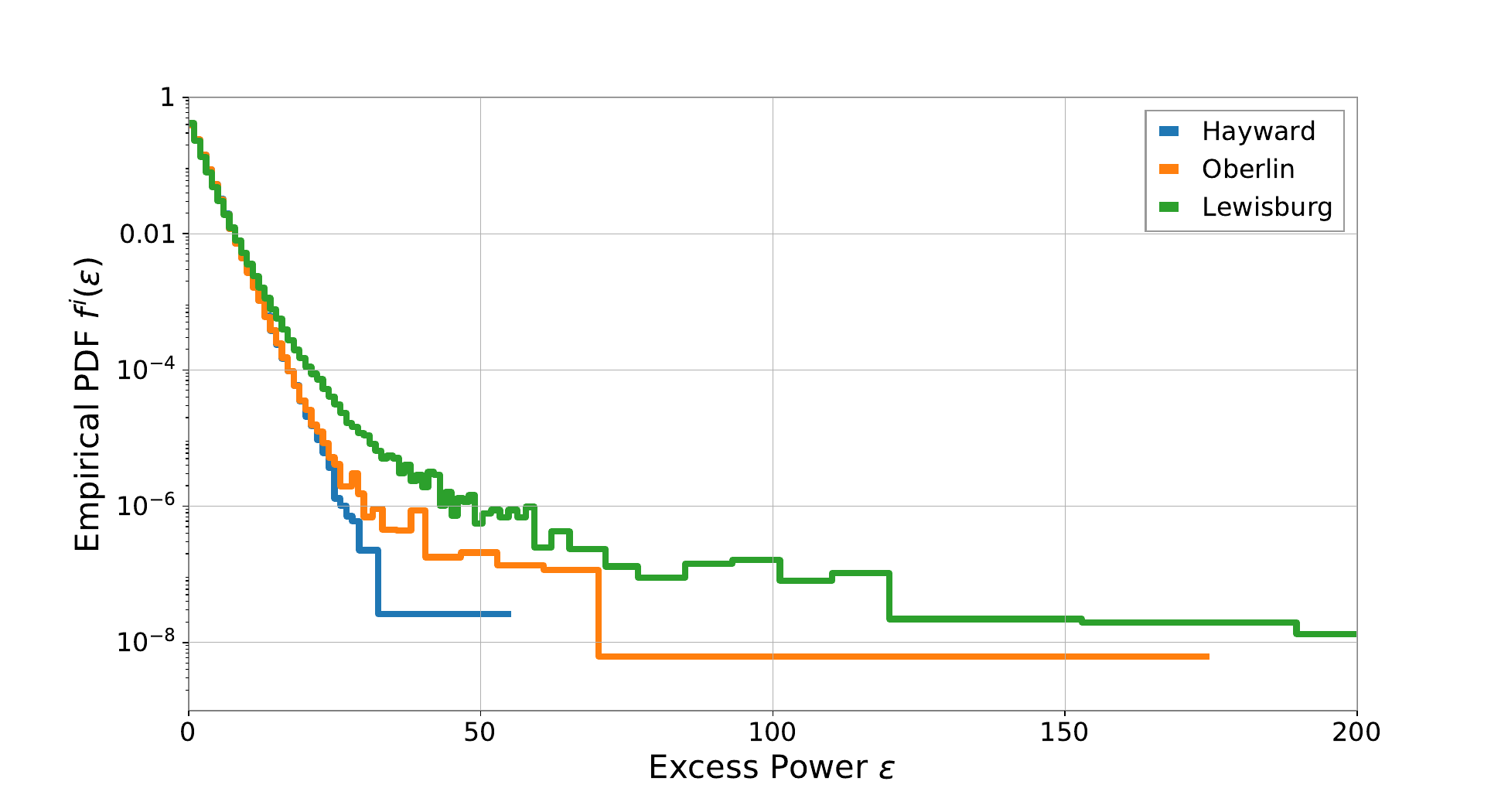}
    \caption{Empirical EP PDFs $f^i(\varepsilon)$ of three stations with tile dimension $5\,\text{s}\times 0.2\,\text{Hz}$. The PDFs serve as a look-up table to convert the EP statistic $\varepsilon$ to a probability density $\rho$. The significantly differing behavior at large $\varepsilon$ highlights the need to construct empirical PDFs to compare stations.}
    \label{fig:EP pdf}
\end{figure}

To form the empirical PDFs, we histogram the background power data with bin widths that monotonically increase as the probability density decreases. The width of the bin for the highest power, which serves as a conservative estimate of the probability density for rare tiles, is set to contain the largest of either the maximum background or foreground tile.  We use the empirical PDF as a look-up table to estimate the probability densities $\rho^i_{jk} = f^i(\varepsilon^i_{jk})$, for every tile. Figure~\ref{fig:EP pdf} depicts PDFs constructed for this search to illustrate the difference between stations.
 
The joint probability density in each tile is found by taking a product of all $N$ active stations' PDFs at each tile
\begin{equation}
\zeta_{jk} = \prod_{i=1}^{N}{\rho^i_{jk}}.
\label{Eq: EPjoint}
\end{equation}
A coincidence threshold $\zeta_{th}$ is determined for each tile size that would pass the loudest (least likely) 1\% of background tiles.  Tiles with joint probability $\zeta_{jk}$ lower than the threshold are passed to the next part of the analysis.

\subsection{Likelihood-ratio test}
\label{sub:Likelihood}

We next employ a LRT to quantify whether candidate tiles are best described by a noise-only or a signal-plus-noise hypothesis. Below, we describe this process for a set of tiles with a single tile dimension.  The same procedure is repeated for each tile dimension.
 
\begin{figure}[h!]
    \centering
    \includegraphics[width = \linewidth]{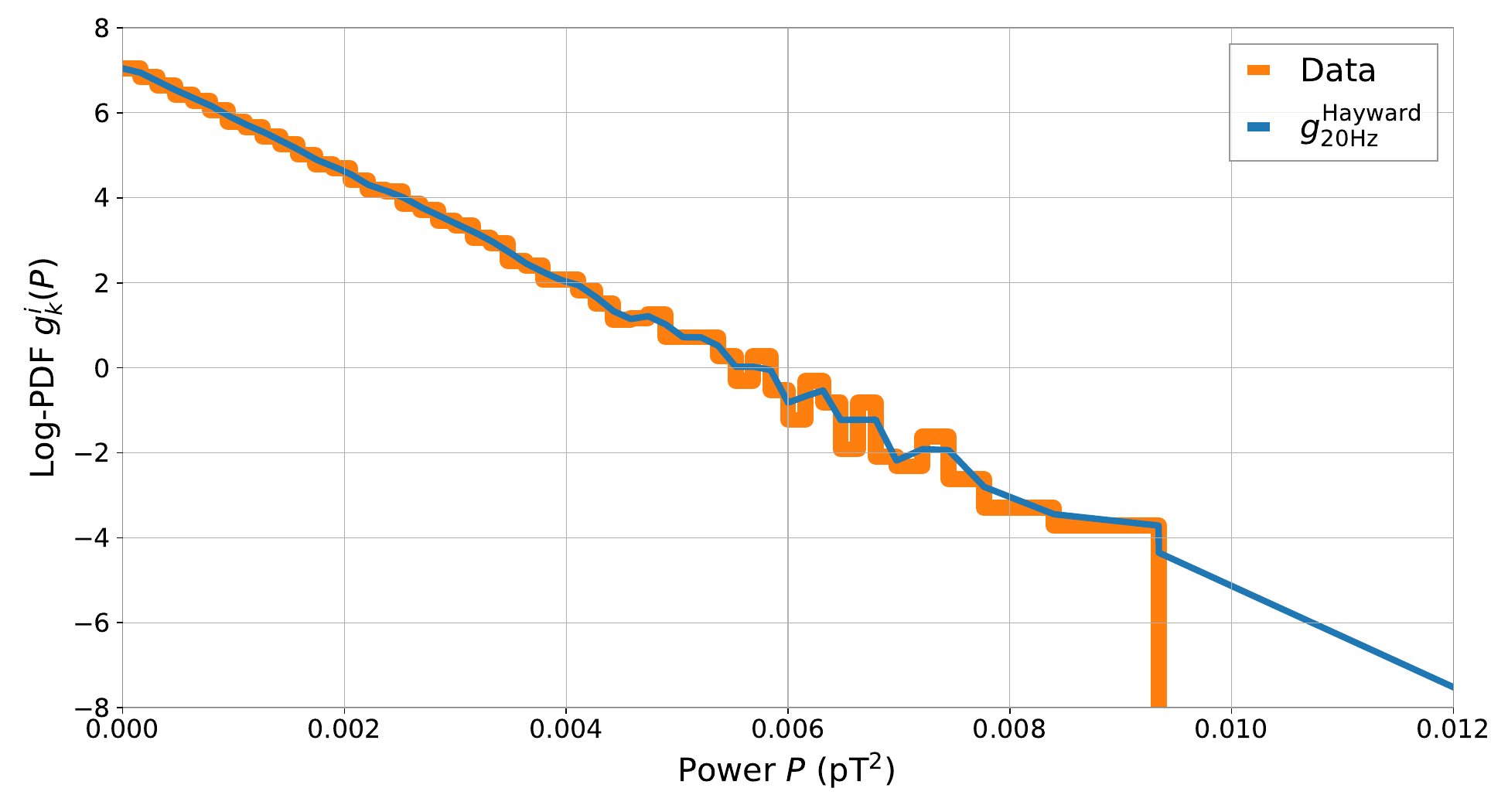}
    \caption{Example of the construction of a noise log-PDF $g^i_k(P)$ from binned data with interpolation and linear extrapolation. Shown is the Hayward station at $20$\,Hz for the tile dimension $2\,\text{s}\times 0.5$\,Hz. $g^i_k(P)$ are constructed for every unique choice of station, tile dimension, and frequency. }
    \label{fig: logpdf}
\end{figure}

First, we characterize the noise behavior of the stations in the absence of a signal.  We construct noise log-PDFs $g^i_k(P)$ of the observed power for each station $i$ at each frequency $k$ from the 24-hour background window. Similar to EP PDFs, we build the noise log-PDFs by interpolating the log of a normalized histogram of the background powers at each frequency, with varying bin widths that contain at least one tile per bin. The tails of the noise log-PDFs are modelled with a linear extrapolation using the background data with the 5\% highest power. Figure~\ref{fig: logpdf} is an example of a log-PDF constructed for a particular station, tile dimension, and frequency. 

The noise log-PDFs $g^i_k(P)$ give the log of the relative probability of detecting a power $P$ in the absence of a signal, i.e. no ELF pseudomagnetic field $\Bp=0$.  In other words, they are the log-likelihood functions $\ell_i(\Bp=0|P)=g^i_k(P)$ for the noise-only hypothesis of a single station given an observed power $P$.  

For the signal hypothesis ($\Bp\neq 0$), we will vary $\Bp$ to find the maximum log-likelihood value, similar to a least squares fit.  The log-likelihood is determined by evaluating the noise log-PDFs at the residual power $\eta^i_{jk}(\Bp)$ given by 
\begin{equation}
    \eta^i_{jk}(\Bp) \equiv \left[\,\left|\tilde{d}^i_{jk}\right| - \left|\tilde{B}^i_{jk}(\Bp)\right|\right]^2 = \left[\sqrt{P^i_{jk}} - \left|\tilde{B}^i_{jk}(\Bp)\right|\right]^2,
\label{Eq: residual}
\end{equation}
where $\tilde{d}^i_{jk}$ is observed signal amplitude, $\tilde{B}^i_{jk}(\Bp)$ is the expected signal amplitude for ELF pseudomagnetic field $\Bp$ [Eq.\,\eqref{Eq:Bj to Bp, more indices}], and $P^i_{jk}$ is the observed power for station $i$.  Absolute values are used since the frequency-dependent phase responses of the individual stations were not calibrated and are unknown.  Note that in the absence of a signal the residual is just the observed power, $\eta^i_{jk}(\Bp=0) = P^i_{jk}$.  

The total log-likelihood $\ell$ for a hypothesis of a signal size $\Bp$, given a set of observed powers at each station $\vec{P}_{jk} \ni P^i_{jk}$, is then given by the sum of the station log-likelihoods 
\begin{equation}
    \ell(\Bp|\vec{P}_{jk}) = \displaystyle\sum_{i=1}^{N} g^i_k \left[\eta^i_{jk}(\Bp)\right].
\label{Eq: Log-likelihood}
\end{equation}
We fit our observed data by varying the ELF pseudomagnetic field $\Bp$ to maximize the log-likelihood $\ell(\Bp|\vec{P})$.  

Finally, the likelihood-ratio test statistic $\lambda$ is typically defined to be twice the log of the ratio of the maximum likelihood for the signal hypothesis to the likelihood of the noise-only hypothesis.   Because we use log-likelihoods this is simply given by the difference:
\begin{equation}
    \lambda_{jk} = 2\left[\max_{\Bp>0}\ell(\Bp|\vec{P}_{jk}) - \ell(\Bp=0|\vec{P}_{jk})\right],
\label{Eq: LRT}
\end{equation}
where a larger value of $\lambda$ means the signal hypothesis is more likely.  

This procedure determines a test statistic for each candidate tile.  However, to determine the statistical significance of a candidate tile, we need to compare the same analysis performed on signal-free data.   This signal-free data is generated by ``time-shifting'' data in the foreground window, i.e. choosing a set of tiles with random times for each station.  After passing the time-shifted data through the EP cut, the same LRT is performed to generate test statistics in the absence of a signal.  

\subsection{Feldman-Cousins confidence belts}
\label{sub:FeldCous}
\begin{figure}[h!]
    \centering    \includegraphics[width =\linewidth]{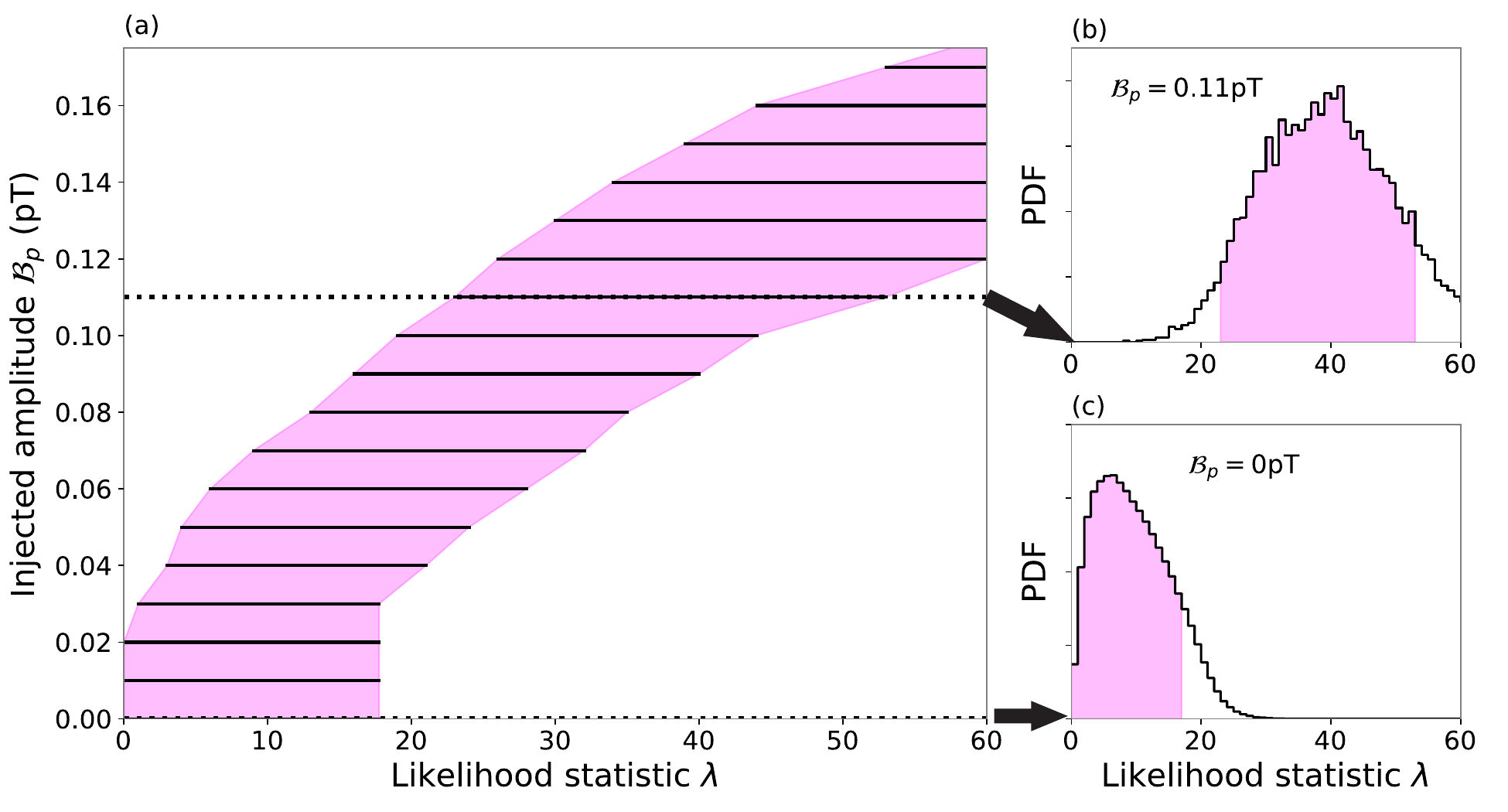}
    \caption{(a) FC plot used to translate $\lambda$ to $\Bp$ in the time-frequency box (see Fig.\,\ref{fig:Adaptive Tiling}) spanning $f_0\in [\text{30,40}]$\,Hz and $\delta t \in [\text{3600,7200}]$\,s with tile dimension $20\,\text{s}\times0.05\,\text{Hz}$. (b) The distribution of $\lambda$ with injected signals of $\Bp=0.11$\,pT.   Shading represents the 90\% acceptance interval from the FC construction that constitutes the horizontal belt. (c) The signal-free distribution of $\lambda$ at $\Bp=0$ from time-shifted data.}
    \label{fig:FCplot}
\end{figure}

We build FC confidence belts~\cite{FeldCous} to determine a confidence interval or an upper limit on the pseudomagnetic ELF amplitude $\Bp$ for a given LRT statistic $\lambda$.  The confidence belts are formed from the distributions of $\lambda$ determined from LRTs with signals of different amplitudes $\Bp$ injected into time-shifted foreground data. For $\Bp=0$, we use the same distributions from the previous section that were determined from the randomly time-shifted data.  As described in Eq.\,\eqref{Eq:DispersionModel}, the signal model is a sine-Gaussian chirp and we inject signals of the form
\begin{equation}
\psi^i(t) = B^i(t|\Bp) e^{-4\ln{2} \frac{t^2}{{\Delta t}^2}} \cos{\left[\left(\omega_0 - \frac{\Delta \omega}{2\Delta t}t\right)t\right]}.
\label{Eq: injection}
\end{equation}
The observed amplitude $B^i(t|\Bp)$ for station $i$ is determined from $\Bp$ and the station sensitivity as described by Eq.\,\eqref{Eq:Bi to Bp}. The chirp rate $\Delta\omega/\Delta\,t$ is chosen with Eq.\,\eqref{Eq:tiledims} for each box in Fig.\,\ref{fig:Adaptive Tiling}. The Gaussian envelope is set to have a full-width half-maximum equal to the tile dimension $\Delta t$.  We inject 1000 chirps for each amplitude $\Bp$ and after performing LRT, we use the resulting distributions of the test statistic $\lambda$ to form the confidence belts in the FC plot as shown in Fig.\,\ref{fig:FCplot}.  We construct a FC plot for every time-frequency box in Fig.\,\ref{fig:Adaptive Tiling}. 

For a candidate tile, a confidence interval for the pseudomagnetic field $\Bp$ can be determined from the intersection of a vertical line at the tile's $\lambda$ with the confidence belts.  If there are no statistically significant tiles in a box the FC belts are used to place an upper limit on $\Bp$.

\subsection{Event significance}
\label{sub:Significance}

To determine the significance of a candidate tile, we first calculate the false positive ratio (FPR), which we define as the number of candidate tiles above a threshold test statistic in a signal-free dataset per total candidate tiles.  The FPR in the $n$th box (see Fig.\,\ref{fig:Adaptive Tiling}) for a given threshold value $\lambda^{th}$ is related to the Cumulative Distribution Function CDF$_n^\lambda$ of the $\lambda$ values of the time-shifted data in box $n$ by  
\begin{equation}
\text{FPR}_n(\lambda^{th}) = \frac{\text{tiles in box}\,n\,\text{of time-shifted data with}\,\lambda>\lambda^{th}}{\text{tiles in box } n\,\text{of time-shifted data}}=1 - \text{CDF}_n^\lambda(\lambda=\lambda^{th}).
\label{Eq:FPR}
\end{equation}
Using many time-shifts allows an estimate of the $\text{FPR}_n$ even for signal thresholds higher than those observed in the foreground.

The FPR of a candidate tile can then be used to calculate the expected mean number of false positives $\mu$ in a dataset.
For the foreground data 

\begin{equation}
    \mu_n(\lambda) = \text{FPR}_n(\lambda)\times N_n^{fg}\,,
    \label{Eq:box fp}
\end{equation}
where $N_n^{fg}$ is the number of foreground tiles in box $n$.   For $\mu\ll1$, the false positive probability (FPP), i.e. the probability to observe one or more false positives, is approximately the mean value\footnote{The probability of getting $k$ events for a Poissonian process with a mean value $\mu$ over an observation window is $\text{Pr}(k)=\mu^k e^{-\mu}/k!$.  For rare events, the probability of getting one or more events in the limit of $\mu\ll1$ is $\text{Pr}(k\ge1)=1-\text{Pr}(k=0)\approx 1-e^{-\mu}\approx\mu.$}, 
\begin{equation}
    \text{FPP}_n(\lambda)\approx \mu_n(\lambda) = \text{FPR}_n(\lambda)\times N^{fg}_n\,.
    \label{eq:FPP_box}
\end{equation}  
Across the entire dataset the $\text{FPP}_{tot}$ for a given FPR is then
\begin{equation}
    \text{FPP}_{tot}\approx \mu_{tot} = \text{FPR}_n(\lambda) \times N^{fg}_{tot}\,,
    \label{eq:FPP_tot}
\end{equation}
where $N^{fg}_{tot}$ is the total number of all foreground tiles.

Using the total number of tiles is equivalent to correcting for the ``look-elsewhere'' effect with a trials factor of $N^{fg}_{tot}/N^{fg}_n$ \cite{lyons2008}.  If the foreground tiles were evenly distributed amongst the boxes (which in our case have very different areas in time-frequency space), this trials factor would be equal to the number of boxes.

The significance of an event detection can be evaluated from its FPP.  For example, a $3\sigma$ detection corresponds to a FPP of $2.7\times 10^{-3}$.  If no significant events are detected, we determine a threshold test statistic for each box such that $\mu_{tot} \leq 1$ for our entire foreground dataset.  

In either case, the FC belts can be used to translate the test statistic into a pseudomagnetic field which can then be converted to a combination of the possible ELF energy scales $f^{l(q)}$, initial energy release $\Delta E$, and initial pulse duration $\tau_0$ as described in Sec.~\ref{sub: detect}. 

\section{Search Target and Results}\label{sec:results}
We now apply this analysis method to the BBH merger GW200311\_115853 detected by LIGO/Virgo on March 11th, 2020 at 11:58:53.398 UTC \cite{PhysRevX.13.041039, PhysRevX.13.041039, GCN_BBH}. The luminosity distance is estimated to be $D_L = 1.17^{+0.28}_{-0.40}$\,Gpc at right ascension 00h08m and declination -07d27m with a sky localization of 35\,deg$^2$. The total mass before the merger was $61.9^{+5.3}_{-4.2}\,M_\odot$ and the final mass was $59.0^{+4.8}_{-3.9}\,M_\odot$. 

\subsection{Foreground analysis}\label{sub:foreground}
GNOME had 5 active stations with quality data at the time of and after the merger that are used in this analysis: Hayward, Lewisburg, Los Angeles, Moxa, and Oberlin. The 10-hour foreground window searched by the analysis is from March 11th, 2020 at 11:58:53 UTC to 21:58:53 UTC and the 24-hour background window is from March 11th, 2020 at 21:58:53 UTC to March 12th, 2020 at 21:58:53 UTC. The background window was taken after the foreground window to maximize the available data from the 5 active stations. 

\begin{figure}[h!]
    \centering
    \includegraphics[width =\linewidth]{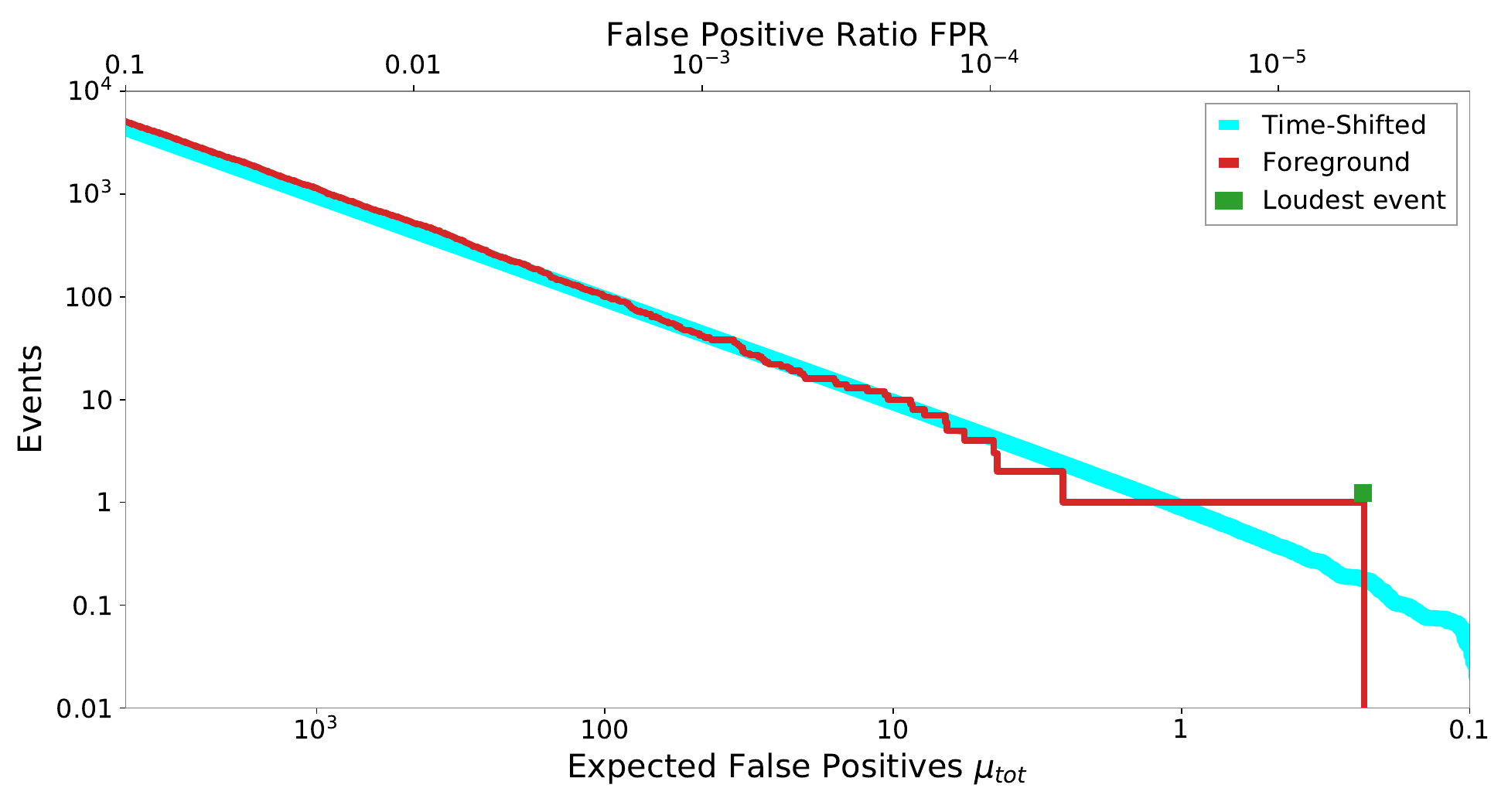}
    \caption{The cumulative number of events above a given FPR [Eq.\,\eqref{Eq:FPR}] for both the time-shifted and foreground data. The mean number of false positives expected for each event is calculated with Eq.\,\eqref{eq:FPP_tot}. The cumulative events for the time-shifted data was divided by the number of time-shifts to represent the average number of false positives expected for the foreground window. The tile with the smallest FPR, indicated by the green square, occurs with an FPP of 0.21 which corresponds to 1.3$\sigma$.}
    \label{fig:foreground}
\end{figure}

Figure~\ref{fig:foreground} shows the cumulative number of foreground events above a given FPP versus the mean number of expected events.   The foreground and time-shifted data are largely consistent.  The most significant event 
 had an expected mean of $\mu_{tot} =0.23$ or FPP$=0.21$, which would correspond to a detection with a significance of only $1.3\sigma$. While this is not significant, we calculate the corresponding ELF masses and energy scales as an example. This tile occurred at a delay time of 15920\,s and a central frequency of 95.8\,Hz. Through the FC method, we obtain a 90\%-confidence upper limit of $\Bp < 0.08$\,pT. For the linear interaction, this corresponds to a mass of  $m = 2.1\times 10^{-19}$\,eV [Eq.\,\eqref{Eq: elf mass}] and linear energy scale of $f_l=2.7$\,GeV at $\tau_0=1$\,s and $\Delta E = 1\,M_\odot c^2$ [Eq.\,\eqref{Eq:fl_lim}]. For the quadratic interaction, the mass is $m=1.0\times 10^{-19}$\,eV and the quadratic energy scale is $f_q=6.7\times10^{-3}$\,GeV at $\tau_0=1$\,s and $\Delta E = 1\,M_\odot c^2$ [Eq.\,\eqref{Eq:fq_lim}].

\subsection{Limits}\label{sub:limit}

\begin{figure}[h!]
    \centering
    \includegraphics[width = \linewidth]{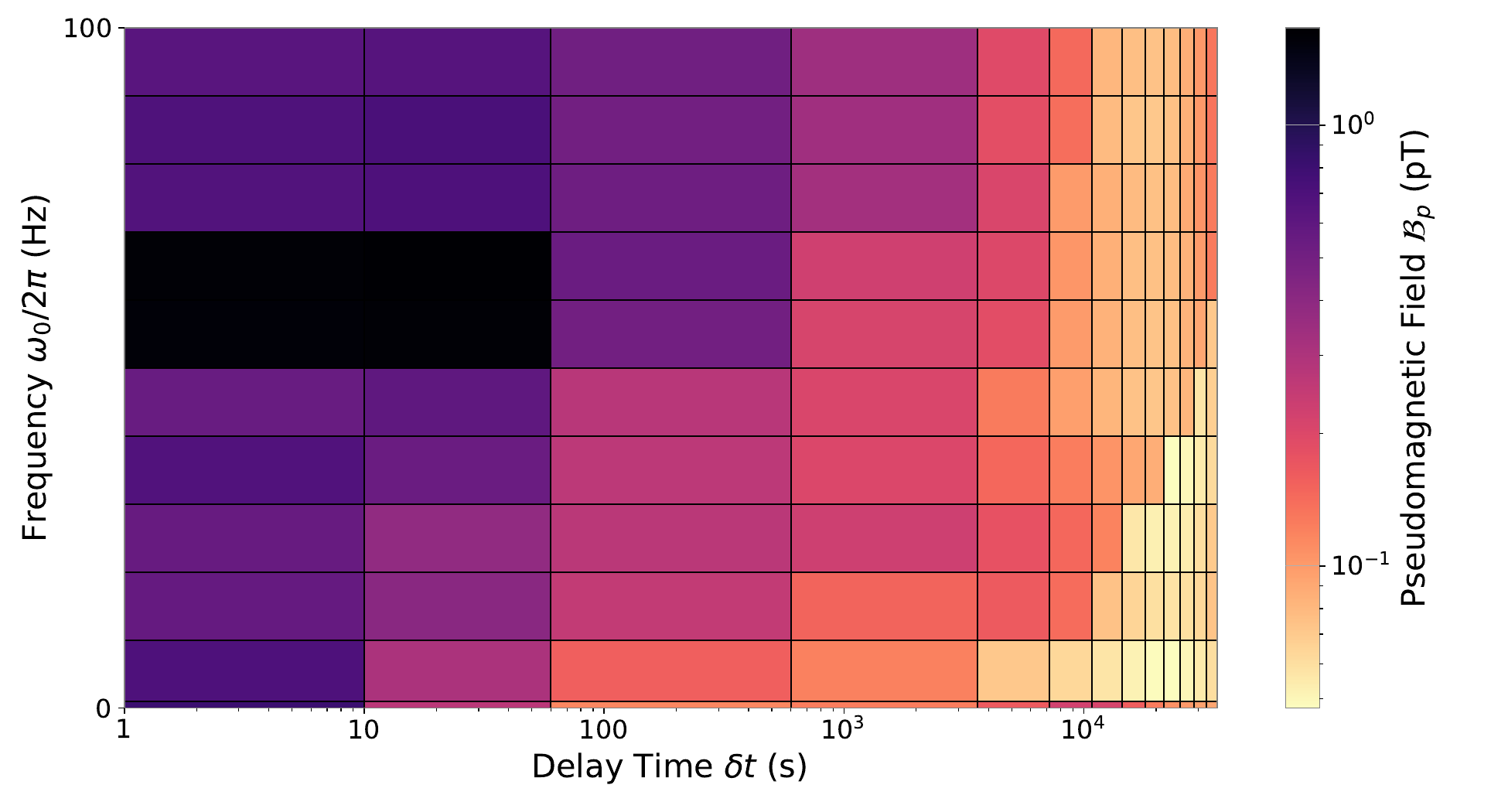}
    \caption{Values of 90\%-confidence upper limits of $\Bp$ at $\lambda^{th}_n$ for each time-frequency box shown in Fig.\,\ref{fig:Adaptive Tiling}.  The limits on $\Bp$ are obtained by converting the threshold to a pseudomagnetic-field amplitude with the FC plot for each box. The network sensitivity strongly depends on the tile dimension deployed with longer time-averaged / narrower bandwidth tiles boasting higher sensitivity. The directional sensitivity of the network contributes to the increase in network sensitivity in the delay times between $\sim 1\times10^4 - 3\times10^4$\,s.}
    \label{fig:bp limit}
\end{figure}

With no significant events detected, we place exclusion limits on the energy scales of the linear and quadratic ELF $\phi$-spin interaction. We determine a threshold $\lambda^{th}_n$ for each box $n$ corresponding to an expected value of one false positive over the dataset, i.e. $\text{FPR}_n(\lambda^{th}_n) = 1 / N^{fg}_{tot}$ [Eq.\,\eqref{eq:FPP_tot}].  These are then translated into a 90\%-confidence upper limit on $\Bp$ with the FC belts constructed for each box which are displayed in Fig.\,\ref{fig:bp limit}.

We determine the ELF mass range probed at each box from Eq.\,\eqref{App:Eq:Lag}. The ELF mass 

\begin{equation}
mc^2 = \hbar \omega_0 \sqrt{\frac{2 c \delta t}{R}},
\label{Eq: elf mass}
\end{equation}
depends on the central frequency $\omega_0$, the delay time $\delta t$, and the distance of the astrophysical event $R$. The dependence on the distance means that different search targets enable us to investigate different mass ranges, while the frequency and time delay dependence defines the mass band that we search for each target. 

\begin{figure}[h!]
    \centering
    \includegraphics[width =\linewidth]
    {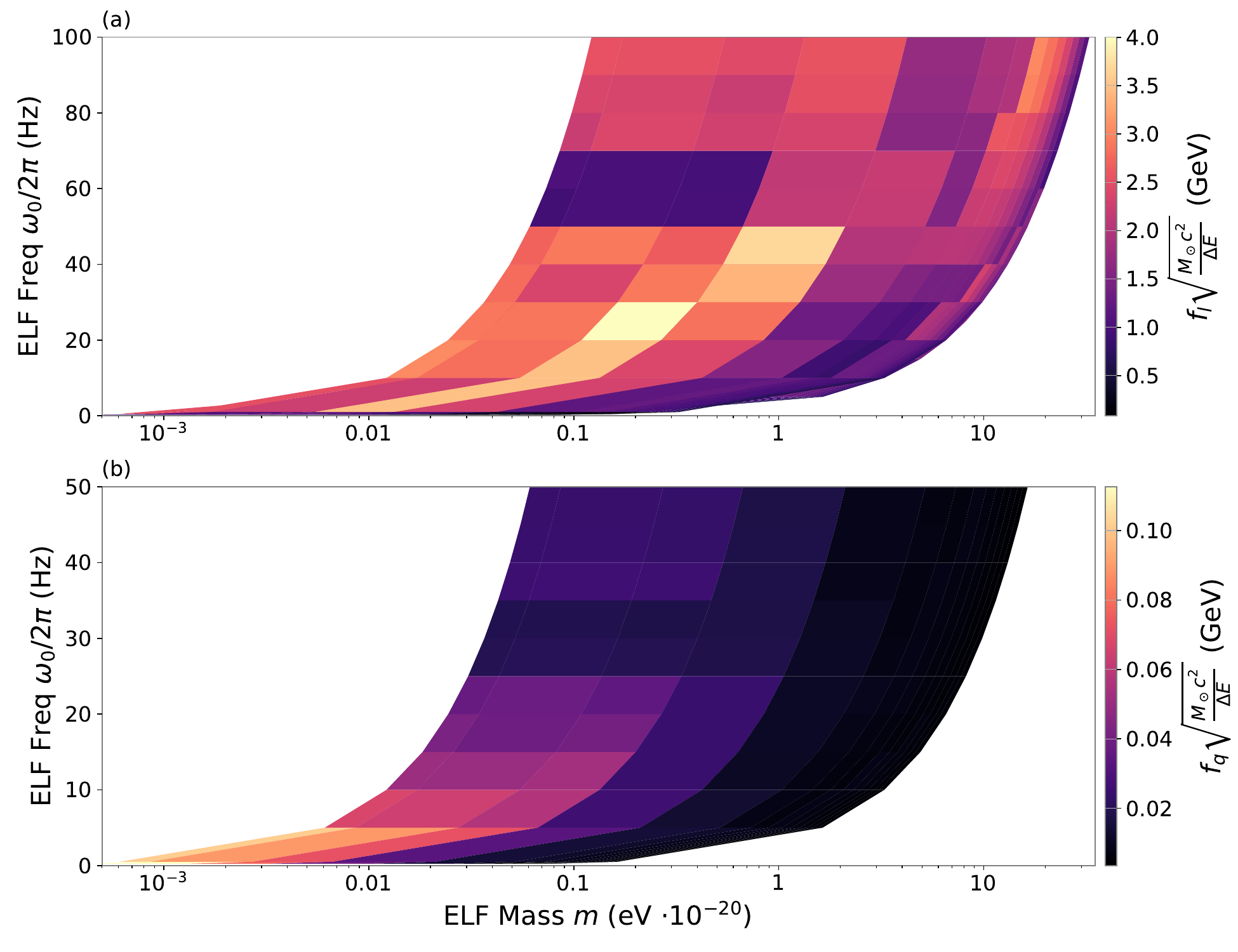}
    \caption{90\% exclusion regions for (a) the linear energy scale $f^l$ [Eq.\,\eqref{Eq:fl_lim}] and (b) the quadratic energy scale $f^q$ [Eq.\,\eqref{Eq:fq_lim}] versus ELF mass [Eq.\,\eqref{Eq: elf mass}] evaluated with $\tau_0 = 1$\,s, $\Delta E = 1\,M_\odot c^2$, and the upper limits on $\Bp$ for each box from Fig.\,\ref{fig:bp limit}. The signal from a quadratic interaction is at twice the central ELF frequency.}
    \label{fig:limits}
\end{figure}

The 90\%-confidence exclusion limits on the energy scale $f^{l(q)}$ [Eqs.\,\eqref{Eq:fl_lim}-\eqref{Eq:fq_lim}] are determined with the 90\%-confidence upper limit on $\Bp$ and the center delay time and frequency of each box. Figure~\ref{fig:limits} displays the exclusion limits calculated for $\tau_0 = 1$\,s and $\Delta E = 1\,M_\odot c^2$. The exclusion limits are the first lab-based constraints placed on the energy scales of the ELF $\phi$-spin interaction. From the energy release, initial burst duration, and distance of a given astrophysical source, we can place a corresponding limit on the ELF parameter space in the case of a non-detection. 

The exclusion limits can be calculated for other values of $\Delta E$ by simply multiplying our presented results by $\sqrt{\Delta E / M_\odot c^2}$. The exclusion limits depend on $\tau_0$ through Eq.\,\eqref{Eq:taufunction}, which depends on the delay time and frequency, and can be scaled to other values of $\tau_0$ by recalculating $\tau$ at the new value for each box. For all boxes, there is a limit on the initial burst duration, $\tau_0 > 1 / \omega_0$, for the emitted energy spectra to be sufficiently sharp in frequency space \cite{Dailey2021,arakawa2025multimessenger}. 

Finally, we assume the ELF pulse and GW/EM trigger are emitted at the same time, for example due to a bosenova/relativistic axion burst \cite{takahashi2024self,aurrekoetxea2024self,eby2022probing,arakawa2024detection}. If an alternative production theory postulates the time of emission of the ELFs and GW/EM differ then a simple timing offset can be added to the analysis.


These limits allow us to explore new territory in both particle physics and astrophysics. From the perspective of BSM physics, our search probes three key aspects of exotic field models: (1) the existence of axion-like particles (ALPs) or other low-mass bosons, (2) their coupling to Standard Model fermion spins, and (3) their potential production in high-energy astrophysical events. While previous laboratory and astrophysical searches have placed bounds on ALP couplings and relic populations, the scenario of coherent bursts produced in dynamic astrophysical environments remains largely unconstrained. Our exclusion plots (Figs.\,\ref{fig:bp limit} and \ref{fig:limits}) do not yet reach the coupling strengths excluded by other laboratory or astrophysical observations \cite{cong2024spin,Raf99,antypas2022new}, but they begin to constrain the combined parameter space of ELF production and coupling. These constraints provide valuable guidance for theory development in this still-nascent area \cite{arakawa2025multimessenger}, including models of bosonic clouds around compact objects \cite{arvanitaki2011exploring,baryakhtar2021black} or bosenova collapse mechanisms \cite{arakawa2024detection}. More broadly, ELF detection or exclusion enables physical insight into the dynamics and environments of extreme astrophysical phenomena. For example, coincident ELF signals could reveal new channels of energy release in black hole or neutron star mergers \cite{alvarez2024finalparsec}, or indicate the presence of novel fields involved in supernova explosions \cite{Raf88} or fast radio bursts \cite{Iwa15,Tka15}. Thus, even null results, as presented here, help narrow the range of viable models and open the door to new observational probes of both fundamental physics and the astrophysical processes that produce them.

\subsection{Projected sensitivity to other events}

We report limits from a single search target but investigating additional events will increase the coverage of ELF parameter space. Both the range of ELF masses [Eq.\,\eqref{Eq: elf mass}] and energy scales [Eqs.\,\eqref{Eq:fl_lim}-\eqref{Eq:fq_lim}] that the network is sensitive to depend on the distance to the astrophysical event.

\begin{figure}[h!]
    \centering
    \includegraphics[width =\linewidth]{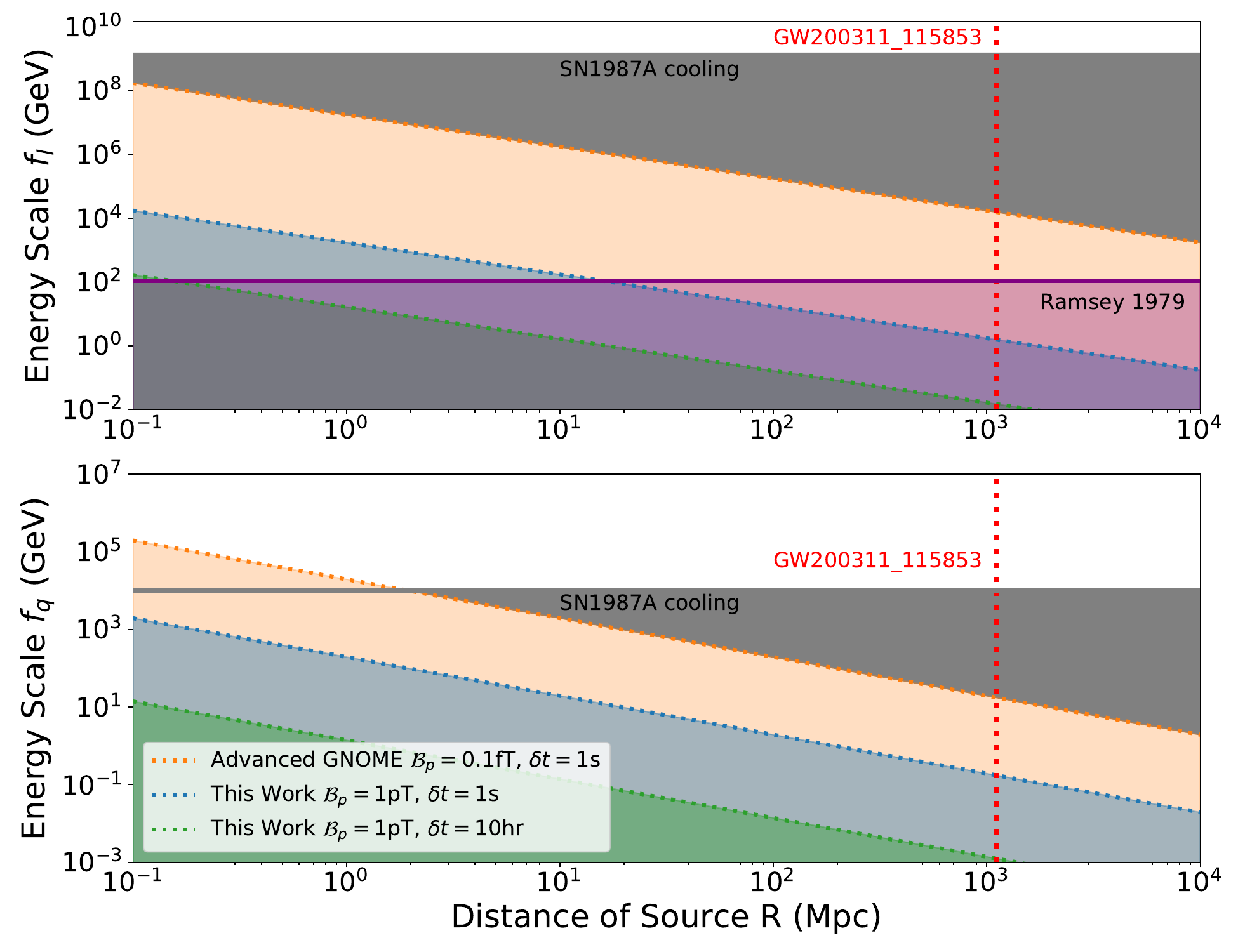}
    \caption{Network sensitivity of this work and projected sensitivity of Advanced GNOME. The curves are Eqs.\,\eqref{Eq:fl_lim} and~\eqref{Eq:fq_lim} calculated with $\tau_0=1$\,s, $\omega_0 / 2\pi = 1$\,Hz, and $\Delta E = 1\,M_\odot c^2$ for different values of $\Bp$ and $\delta t$. The gray area represents the astrophysical exclusion limits on $f_l$ and $f_q$ which are $1.5\times10^9$\,GeV \cite{Lella_2024} and $1\times10^4$\,GeV \cite{pospelov2013detecting}, respectively. The purple area represents the parameter space ruled out by laboratory measurements of proton-spin couplings~\cite{ramsey1979tensor,cong2024spin}. The vertical dotted line indicates the distance to the BBH merger GW200311\_115853 probed in this paper.}
    \label{fig:proj sens}
\end{figure}

The GNOME collaboration is upgrading station sensors to co-magnetometers for the ``Advanced GNOME'' network \cite{Afach2023} that boast orders of magnitude higher sensitivity to ELF signals \cite{PhysRevResearch.6.013339_comag}. Figure~\ref{fig:proj sens} displays the projected sensitivity of this analysis to the ELF energy scales across a range of astrophysical distances. The curves are calculated for the approximate network sensitivity in this work and the projected network sensitivity of Advanced GNOME. The associated energy scales for linear and quadratic ELF interactions [Eqs.\,\eqref{Eq:fl_lim}-\eqref{Eq:fq_lim}] are evaluated at $\tau_0=1$\,s, $\omega_0 / 2\pi = 1$\,Hz, and $\Delta E = 1\,M_\odot c^2$ over a range of astrophysical distances. Astrophysical limits based on the rate of SN1987A cooling on $f_l$ and $f_q$ are $1.5\times10^9$\,GeV \cite{Lella_2024} and $1\times10^4$\,GeV \cite{pospelov2013detecting}, respectively. GNOME could search parameter space beyond the indirect astrophysical limits with the next generation of co-magnetometer sensors.

Expanding the number of detectors in the network is another route to increasing sensitivity and false positive rejection. The question of the optimal configuration of the network was explored in Ref.~\cite{smiga2022assessing}. Each detector makes an independent measurement of $\Bp$ based on the projection of $\Bp$ on its sensitive axis. As three stations are required to define the projection of $\Bp$ on the network, we have that to first order, the signal-to-noise ratio of the detection LRT statistic will scale as $\sqrt{N}$ for $N>3$ under the assumption of equivalent sensitivity for each detector. Corrections to this scaling -- accounted for in this analysis -- arise due to the particular target considered and the relative detector noise. Fig.\,\ref{fig:bp limit} shows an effective true positive network sensitivity in $pT$ for GW200311\_115853 after folding in the noise characteristics, and orientation of the 5 stations used in this search. As noted in Ref.\,\cite{pospelov2013detecting}, increasing the number of detectors can give faster than $\sqrt{N}$ improvement for false positive rejection in the case where the background is dominated by rare non-Gaussian noise spikes.

\section{Conclusion}
Using the Global Network of Optical Magnetometers for Exotic physics (GNOME), we have performed a multi-messenger astronomy search for ultralight, scalar BSM fields. This work considered the emission of an exotic low-mass field (ELF) burst produced during the binary black hole merger GW200311 from which gravitational waves were detected by LIGO/Virgo~\cite{GCN_BBH}. No significant events were detected by GNOME. We consequently place the first lab-based constraints on combinations of ELF production parameters, namely the energy scales of the interaction between the ELF and proton spins, the energy released from the astrophysical event in the form of an ELF burst, and the initial ELF-burst duration.

In the future, this analysis method will be applied to existing GNOME datasets \cite{Afach2023} which overlap with other gravitational-wave events detected by LIGO/Virgo~\cite{PhysRevX.13.041039} and dozens of fast radio bursts~\cite{frbcollaboration2025cataloglocaluniversefast}. This will expand the mass range explored and closer events will allow higher energy scales to be probed. The collaboration is further upgrading to self-compensating noble-gas-alkali-metal comagnetometers for implementation in the Advanced GNOME experiment \cite{klinger2023optimization,PhysRevResearch.6.013339_comag}. These improved sensors will be able to probe the coupling of exotic fields to protons and neutrons with higher sensitivity than our current sensors which can only probe the coupling to protons. Additionally, the response to magnetic and non-magnetic signals in the comagnetometers can be discriminated, allowing for a reduction of the false positive rate \cite{Afach2023}.  As an extension of this work, we plan to generalize this technique to an all-sky search to hunt for potential signals without a known trigger.  In the event of a detection, the network can truly act like a telescope with the ability to localize the signal origin. Any ELF detection would be indicative of new physics -- in particular, an ELF observation would not only be a discovery of a BSM particle, but would give insight into as yet unknown production mechanisms in an extremely high energy and strongly curved spacetime regime.

The development of quantum sensors has led to a variety of measurements with exquisite sensitivities in individual laboratories \cite{Ye2024}.  Using GPS, it is now straightforward for labs to accurately time stamp data -- enabling new networks of quantum sensors.  To analyze data from a different quantum sensor network using the algorithm described in this paper, one would primarily have to change the function that describes the network response to an ELF, e.g. Eq.~\eqref{Eq:Bj to Bp, more indices}, for these other sensors and account for differing sampling rates and bandwidths. We propose that the analysis framework presented here can serve as a template that can be used for a variety of networks searching for BSM physics. Multimessenger astronomy with quantum sensors will inform investigations into astrophysical production mechanisms of BSM fields, their propagation, and their interactions with matter -- opening new windows with which to view the universe.

\section*{Acknowledgements}

This work was supported by the U.S. National Science Foundation under grants PHY-1707875, PHY-1707803, PHY-2110370,  PHY-2110388, and PHY-2207546; by the German Research Foundation (DFG) under grant no. 439720477 and within the German Excellence Strategy (Project ID 39083149); by work from COST Action COSMIC WISPers CA21106, supported by COST (European Cooperation in Science and Technology)”;
ZDG acknowledges institutional funding provided by the Institute of Physics Belgrade through a grant by the Ministry of Science, Technological Development and Innovations of the Republic of Serbia.
The work of SP is supported by the National Science Centre of Poland within the Opus program (grant 2019/34/E/ST2/00440) and GL acknowledges the support of the Excellence Initiative -- Research University of the Jagiellonian University in Krak\'ow. The authors thank Yuanhong Wang for pointing out an error in our calculation in the preprint version of this article.

\bibliographystyle{unsrt}
\bibliography{references,ELFbib}

\end{document}